# Ion Concentration and Voltage Imaging with Fluorescent Nanodiamonds

Patrick Voorhoeve[1], Hiroshi Abe[2], Takeshi Ohshima[2,3], Qiang Sun[4], Anita Quigley[1,5,6], Rob Kapsa[1,5,6], Nikolai Dontschuk[7], Philipp Reineck[4*]

[1] Biomedical Engineering, School of Engineering, RMIT University, Melbourne, VIC, Australia.
[2] National Institutes for Quantum Science and Technology (QST), Takasaki, Gunma, 370-1292, Japan.
[3] Department of Materials Science, Tohoku University, Aoba, Sendai, Miyagi 980-8579, Japan
[4] School of Science, RMIT University, Melbourne, VIC 3001, Australia
[5] Aikenhead Centre for Medical Discovery, 27 Victoria Parade, Fitzroy, Melbourne, VIC 3065, Australia.
[6] Clinical Neurosciences, St Vincent's Hospital Melbourne, Fitzroy, Melbourne, VIC 3065, Australia
[7] School of Physics, University of Melbourne, Parkville, Victoria 3010, Australia.
7

*email: philipp.reineck@rmit.edu.au

The nitrogen-vacancy (NV) center in diamond exists in different charge states with distinct photoluminescence properties, which are sensitive to the nanoscale electrochemical environment. Hence, the NV charge state is emerging as a powerful all-optical platform for nanoscale sensing and imaging. Although significant progress has been made in engineering near-surface NV centers in bulk diamond, controlling the NV charge state in fluorescent nanodiamonds (FNDs) has proven challenging, limiting the sensitivity and reliability of FND-based charge state sensing. Here, we demonstrate reliable, reversible switching between the fluorescent $NV^0$ and non-fluorescent $NV^+$ charge states in sub-30 nm FNDs via surface oxidation and hydrogenation, respectively, for single particles and particle powder. In aqueous electrochemical cells, we demonstrate voltage and ion concentration imaging based on the NV charge state in self-assembled FND layers on transparent substrates. Applied voltages reliably modulate the FND PL with a sensitivity of up to 16 mV $Hz^{-1/2}$. Importantly, FND PL is also modulated by local changes in salt concentration with a sensitivity of up to 1.8% per millimolar NaCl, enabling all-optical imaging of ion concentration gradients at the microscale. Our results represent a significant step toward realizing fast, stable, and scalable nanoscale charge- and voltage-imaging technologies with sub-micrometer spatial resolution.

**1. Introduction**. The nitrogen-vacancy (NV) center in diamond can exist in three different charge states: $NV^-$, $NV^0$, and $NV^+$. Only $NV^-$ and $NV^0$ are photoluminescent, and only the $NV^-$ has an optically addressable spin and can be used for spin-based quantum sensing. Spin-based quantum sensing of temperature[1–4], magnetic fields[5–8], and pH[9,10] using $NV^-$ centers in nanodiamonds has been studied for a wide range of applications, many of which are in biology[11–13]. Spin-based electric-field sensing has been demonstrated in bulk diamond[14,15] but not in nanodiamonds. However, spin-based sensing generally requires microwave fields for spin control, which are strongly absorbed by many liquids, including water, thereby limiting its applicability, particularly in biology and chemistry. NV charge state sensing is emerging as a powerful all-optical alternative[16–19]. Charge state sensing is based on the NV's ability to reversibly accept or donate electrons in response to changes in its local electrochemical environment, thereby altering NV PL color, intensity, or both. Here, NV surface proximity and control of the diamond surface chemistry are paramount to enable and control these electron transfer processes via surface-induced band bending[20,21]. Oxidized and hydrogenated diamond surfaces exhibit positive and negative electron affinities[22], respectively, and have been studied extensively in this context. In an aqueous environment, oxygen-containing diamond surface groups stabilize $NV^-$ by limiting charge transfer from NV centers to surface states. Hydrogen-terminated diamond surfaces exposed to the atmosphere form a layer of adsorbates that accept electrons from the diamond valence band, creating an electron-depletion layer (and a conductive two-dimensional hole gas) near the surface. NVs within this layer lose electrons and transition to $NV^0$ or $NV^+$ [23,24]. In bulk diamond, this effect has allowed shallow NV centers within a few nanometers from a partially hydrogenated surface to enable fast and sensitive voltage imaging in an electrochemical cell[18].

While near-surface NVs must be carefully engineered in bulk diamond sensing chips, NV centers in small sub-30 nm fluorescent nanodiamonds (FNDs) are inevitably near the surface and can be produced at scale. Efficient NV charge state modulation of hydrogenated FNDs below 30 nm in size in direct contact with an electrode has been demonstrated in an aqueous electrochemical cell[17]. However, since this first report, few studies have made significant progress in FND-based charge state sensing. The NV charge state in FNDs was used to detect charged molecules[25] and to measure pH changes[26]. Menon et al recently reported NV charge state modulation in nanodiamonds using cyclic voltammetry in the presence of a redox-active protein[27]. We have recently demonstrated that the NV PL from FNDs embedded in a capacitor device, but not in contact with an electrode, is modulated by electric fields[28].

Here, we demonstrate reliable, reversible switching between the fluorescent $NV^0$ and non-fluorescent $NV^+$ charge states in sub-30 nm FNDs via surface oxidation and hydrogenation, respectively. Employing a recently developed color center preserving hydrogenation technique[29], we fabricated hydrogenated FNDs (FND-Hyd), in which most NV centers are in the $NV^+$ charge state, and show that these can be electrostatically self-assembled[30] onto glass substrates. In single particle and particle ensemble experiments, we show that the $NV^0$ charge state can be largely recovered via a simple UV-ozone treatment. In an aqueous electrochemical cell, we then study the response of FND-Hyd and FNDs with an oxidized surface (FND-Oxy) to applied voltages. While FND-Oxy shows only a minimal change in PL, the PL of FND-Hyd reliably increases and decreases upon application of positive and negative voltages, respectively, with a sensitivity of up to 16 mV $Hz^{-1/2}$. Finally, we investigate the effect of the local salt concentration on FND-Hyd PL and find that the $NV^0$ PL intensity is also modulated by local changes in NaCl concentration with a sensitivity of up to 1.8% $\Delta PLmM^{-1}$ NaCl.

**2. Results and Discussion**. Figure 1 shows an overview of the FND materials, fabrication techniques, and samples employed in this study. NV proximity to the diamond surface and close control over the diamond surface chemistry are two important prerequisites for efficient charge state sensing. Hence, we focus on sub-30 nm FNDs with either an oxidized (FND-Oxy) or hydrogenated (FND-Hyd) surface (Figure 1a and 1b, respectively). While the oxidized surface has a positive electron affinity (PEA), stabilizing the $NV^0$ charge state in sub-30 nm FNDs, a hydrogenated FND surface has a negative electron affinity (NEA), allowing for electron transfer from near-surface NVs to surface acceptor states. This leads to the formation of a depletion layer, converting near-surface NVs to $NV^+$. Photoluminescence (PL) images of FND-Oxy particles on a solid substrate in air show bright $NV^0$ PL (Figure 1d), which is about an order of magnitude stronger than that of FND-Hyd particles (Figure 1e). In sub-30 nm FNDs, this strong change in NV PL can, in general, be explained with near-surface band bending[24]. The PEA of the FND-Oxy surface leads to a downward bending of the NV charge state transition levels relative to the Fermi level ($E_F$), stabilising the $NV^0$ charge state within this region (Figure 1f). Conversely, the NEA of the FND-Hyd surface induces an upward bending, creating more $NV^+$ near the surface (Figure 1g), with only NV centers in the center of the particle remaining in the $NV^0$ state. The residual $NV^0$ PL from FND-Hyd particles is more sensitive to external stimuli, which will be investigated in Figures 4 and 5.

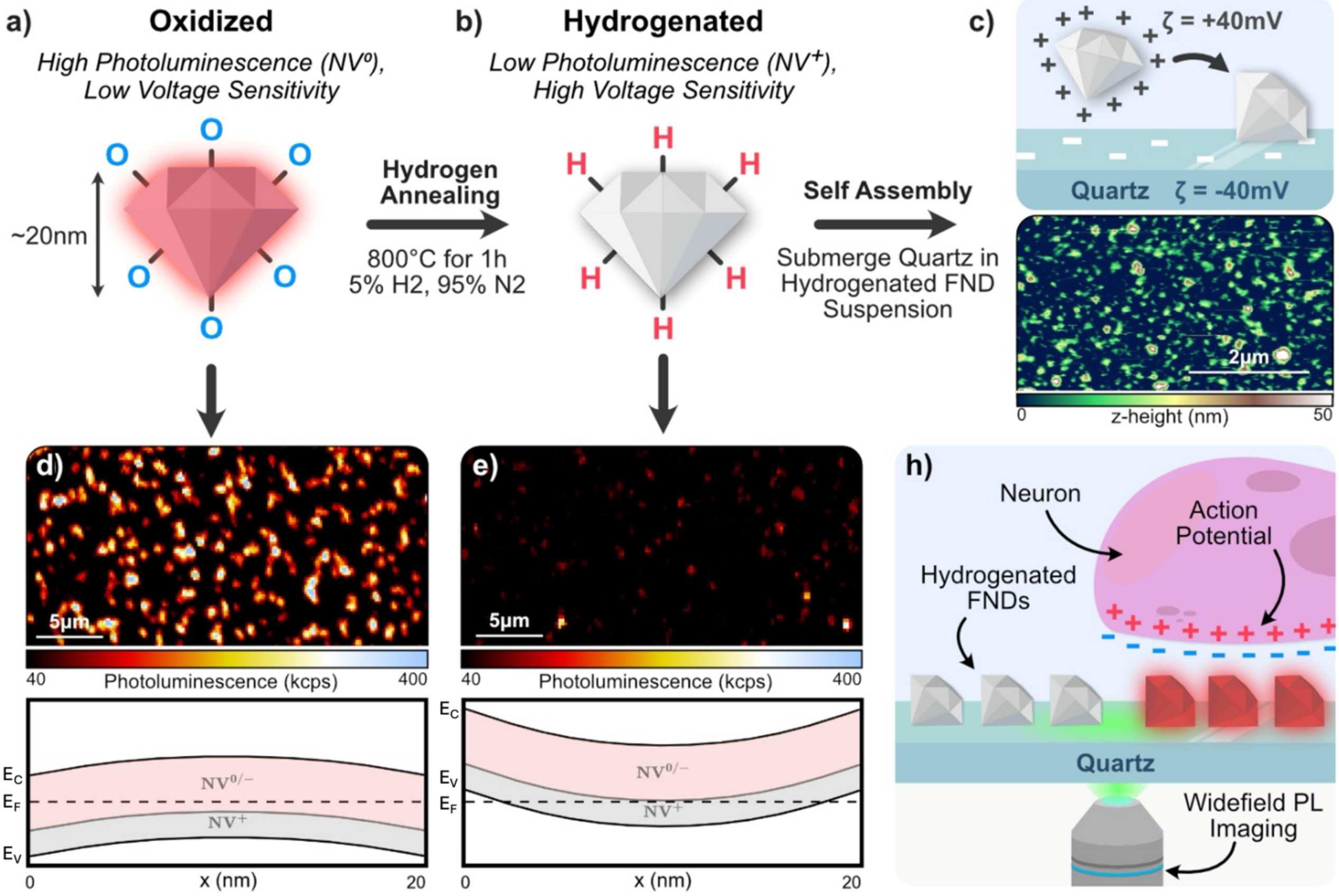

**FIGURE 1** | Overview of the fluorescent nanodiamonds (FNDs) and processing techniques used in this study and their envisaged application. **a-c)** Schematic illustrations of the FND processing **(a,b)** and imaging substrate fabrication process **(c)**. Oxidized FNDs (FND-oxy) show strong $NV^0$ photoluminescence (PL), which is reduced upon hydrogen surface termination (FND-Hyd) via annealing in forming gas due to the creation of $NV^+$. The FND-Hyd particles exhibit a positive zeta potential in water at neutral pH and self-assemble onto negatively charged substrates. **d,e)** Confocal PL images of FND-Oxy **(d)** and FND-Hyd **(e)** on a quartz substrate and corresponding energy diagrams illustrating the surface-termination-induced band bending of the NV charge state transition levels for 20 nm FND-Oxy and FND-Hyd. **h)** Illustration of the envisaged application of FND voltage imaging chips. Neurons are grown on a transparent imaging chip coated with FND-Hyd particles. The FND-Hyd PL switches on in response to a neuronal action potential and is very low otherwise.

At the same time, the zeta potential of FND-Hyd particles suspended in water at neutral pH becomes highly positive[31], enabling electrostatic self-assembly of particles onto any negatively charged substrate (Figure 1c), analogous to a process investigated in detail by our team recently[30]. The process is simple and scalable, and only involves immersing a clean substrate in an FND-Hyd suspension (see the Methods section for details). Figure 1 c), bottom, shows an atomic force microscopy image of FND-Hyd particles self-assembled onto a quartz substrate. Eventually, such substrates may enable all-optical neuronal and neural imaging with high spatiotemporal resolution as illustrated in Figure h).

**2.2 Reversible Switching of the NV Charge State in FNDs**. A key prerequisite to achieving this goal is control over the NV charge state via the diamond surface chemistry. Hence, we first investigated whether surface oxidation and hydrogenation can be used to reversibly switch between the $NV^0$ and $NV^+$ charge states. Figure 2a) shows a confocal PL image of commercially available 20 nm FNDs (Adamas Nanotechnologies, USA) with an oxidized surface, electrostatically self-assembled onto a marked quartz substrate and a zoomed-in ~6×6 μm$^2$ image of four regions of interest (ROI, circles). See Experimental Section and Supporting Information (SI) Figures S1 to S3 for details. Figure 2f) shows an AFM image of the zoomed-in region and individual images of the particles in the ROIs. AFM images suggest that the ROIs contain small FND aggregates 100-200 nm in diameter in the x-y imaging plane with z-heights of 25 nm or less. Using 520 nm laser excitation (900 µW, PL collected above 550 nm), these aggregates show PL intensities between 290,000 and 75,000 counts per second (cps), similar to many other particles seen in the larger field of view (FOV) in Figure 2a).

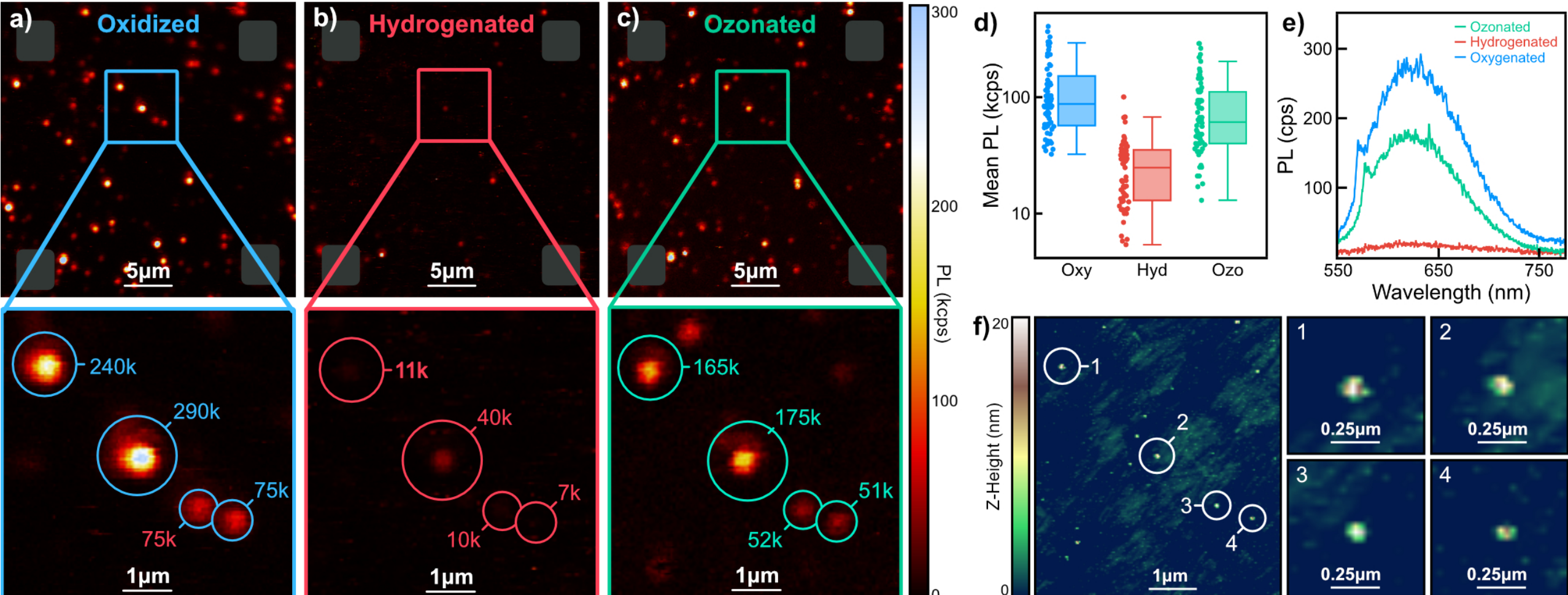

**FIGURE 2** | Reversible switching of the NV charge state in individual small FND aggregates. **a-c)** Confocal PL images of FNDs on a marked substrate. The same FNDs were imaged in their oxidized form **(a)**, after hydrogenation **(b)** and after a UV-ozone treatment **(c)**. The top images show a $25 \times 25\ \mu m^2$ area on the substrate, and the bottom image a zoomed-in image. **d)** Box plot of the PL intensity of 95 FNDs after the different processing steps. Small dots represent individual measurements, the box plot represents the average (horizontal line) and the upper and lower quartiles, and the whiskers the minimum and maximum PL values. **e)** Typical PL spectra of a small FND aggregate after the different processing steps. **f)** AFM z-height images of the zoomed-in image region in the bottom row of panels a-c.

The FNDs were then hydrogenated at 800 °C for 1 h in forming gas (95% N, 5% $H_2$). Figure 2b) shows a PL image of the region in Figure 2a) after hydrogenation, acquired using identical imaging conditions. The images show a significant overall reduction in PL intensity by more than one order of magnitude in some ROIs. We then exposed the FNDs to an ozone atmosphere using a basic benchtop UV-ozone cleaner for 15 minutes. Figure 2c) shows the same sample region after this ozone treatment (FND-Ozo), revealing a significant recovery of PL across the FOV. Qualitatively, this demonstrates that the reduction in PL after hydrogenation is not caused by annealing-induced damage to the FNDs (e.g. via graphitization) but by reversible, surface-chemistry-induced modulation of the NV charge state between the fluorescent $NV^0$ and non-fluorescent $NV^+$ charge state.

To quantify the effect of hydrogenation and ozonation, we investigated the PL of 95 individual small FND aggregates in their oxidized, hydrogenated and ozonated forms. Figure 2d) shows the total PL intensity of all particles (dots) and their mean PL intensities (box plot) for the different processing stages on a logarithmic scale. The box plot shows the mean PL intensity (horizontal line), upper and lower quartiles (box), and minima and maxima (whiskers). While individual particles show a broad range of PL intensities at all processing stages, the average PL intensities change from 87 kcps (FND-oxy) to 25 kcps (FND-Hyd), to 61 kcps (FND-Ozo). Figure 2e) shows representative PL spectra from the same ROI for the different processing steps, revealing $NV^0$ PL spectra with the characteristic $NV^0$ zero phonon line at 575 nm for the oxidized and ozonated FNDs, and very low PL for the hydrogenated FNDs. While the $NV^0$ PL doesn't fully recover after the ozone treatment, these experiments demonstrate that hydrogenation via annealing in forming gas, followed by re-oxidation via an ozone treatment, enables the reversible switching between $NV^0$ and $NV^+$ in sub-30 nm FNDs. One possible explanation for the incomplete recovery of $NV^0$ PL after the UV-ozone treatment is that ozone molecules may not be able to access the FND facet in contact with the substrate, whereas much smaller hydrogen molecules can.

**2.2 Scalable Functionalization of ND Powders**. As shown above, processing FNDs dispersed on a solid substrate enables a systematic investigation of the effects of each processing step on the PL properties of single FNDs and small aggregates. However, most applications require the scalable functionalization of FND powders at the milligram scale or above. Furthermore, most substrates do not withstand the harsh conditions during annealing, and directly investigating changes in surface chemistry of individual particles, e.g., via infrared spectroscopy, is challenging. Hence, we next investigated whether the hydrogen annealing approach used above leads to a similarly efficient switching of NV PL in FND powders. This process is scalable, and the resulting particles would enable the fabrication of FND-Hyd layers on the centimeter scale using electrostatic self-assembly[30]. We used FNDs with a nominal particle size of 18 and 120 nm that were irradiated with 2 MeV electrons and annealed in argon to create NV centers and oxidized in air to create an oxidized surface. See Experimental section for details. We chose two particle sizes to investigate the effect of NV surface proximity on the NV charge state switching efficiency. 150 mg of particle powder for each particle size was annealed in forming gas at 800°C for 1 h. Figure 3a) shows Fourier-transform infrared (FTIR) absorption spectra of the particle powders before and after hydrogenation. For both particle sizes, the O–H and C=O absorption peaks at 1626 $cm^{-1}$ and 1779 $cm^{-1}$, respectively, observed in the oxidized particles, decrease significantly upon hydrogenation. At the same time, two peaks at 2832 $cm^{-1}$ and 2932 $cm^{-1}$ are associated with $CH_x$ surface groups that appear after hydrogenation, suggesting a pronounced change in the particles' surface chemistry.

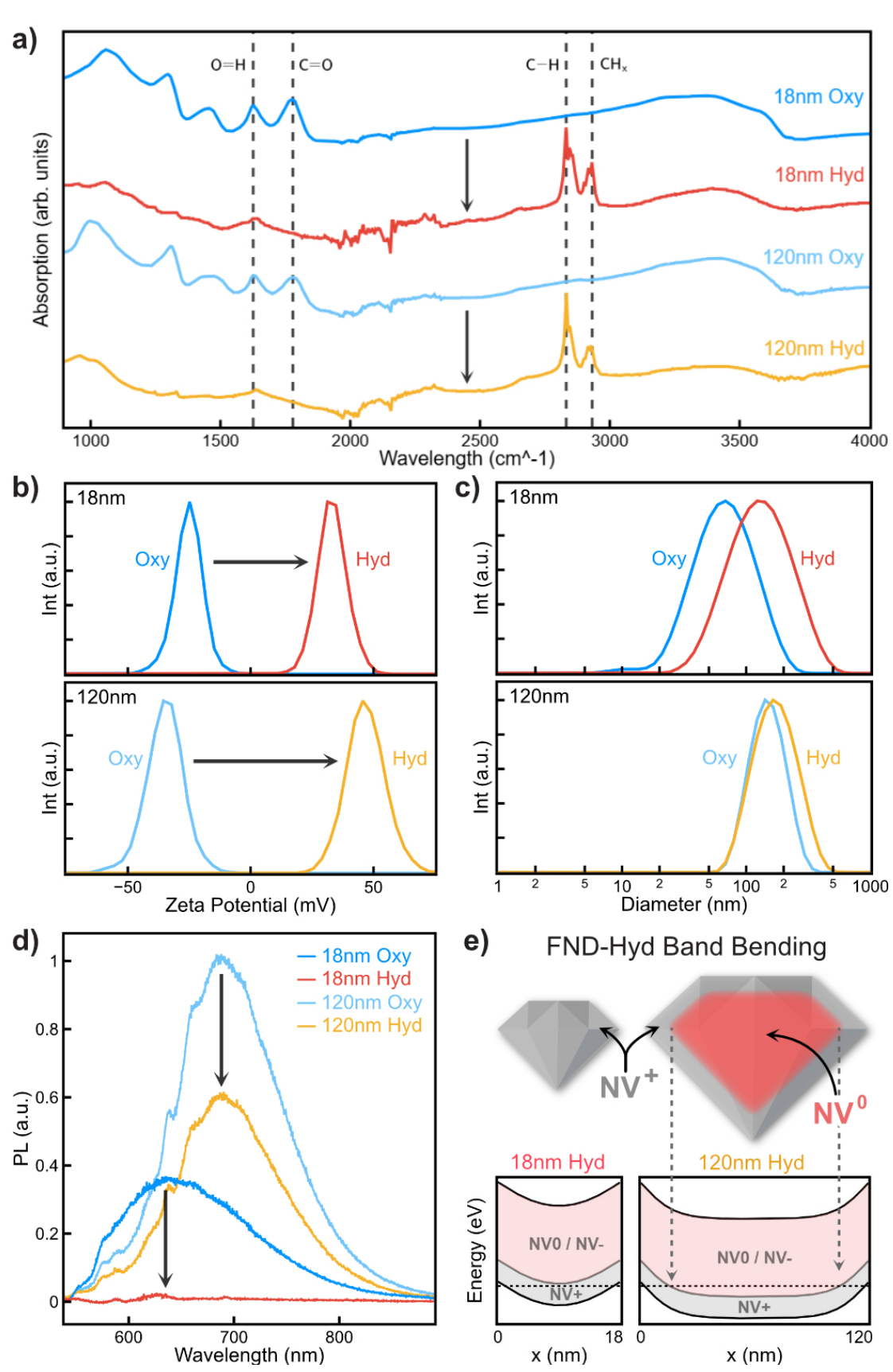

**FIGURE 3** | Characterization of FND powders before and after functionalization. **a)** FTIR spectra of 18 nm and 120 nm FND-Oxy and FND-Hyd powders. **b,c)** Zeta potential distribution **(b)** and dynamic light scattering particle size distributions **(c)** of 18 nm (top) and 120 nm (bottom) FNDs suspended in water before and after hydrogenation. **d)** PL spectra of 18 nm and 120 nm FNDs suspended in water at 1 mg $mL^{-1}$ and 0.1 mg $mL^{-1}$, respectively. **e)** Schematic illustrations of the NV charge state distribution (top) and energy diagrams of the near-surface band bending (bottom) in 18 nm and 120 nm FNDs.

We then suspended the FNDs in deionized water using probe sonication to investigate their colloidal and PL properties. Figure 3b) and 3c) show zeta potential measurements and dynamic light scattering (DLS) particle size distributions, respectively, for all samples. The average zeta potential for 18 and 120 nm FNDs shifts from below -25 mV in their oxidized form to above 25 mV after hydrogenation. At the same time, the DLS particle size distribution shifts to larger particle diameters after hydrogenation. The latter may cause a decrease in particle dispersibility or annealing-induced particle aggregation that cannot be reversed by sonication. The data in Figure 3c) shows unweighted intensity distributions. These suggest typical particle diameters of ~30 nm and ~50 nm for the FND-oxy and FND-Hyd, respectively, with a nominal particle size (provided by the manufacturer) of 18 nm; for the FNDs with a nominal particle size of 120 nm, the particle diameter increases only slightly after hydrogenation, and the nominal particle size agrees with DLS results.

Figure 3d) shows normalized PL spectra of all samples suspended in water (1 mg $mL^{-1}$ for 18 nm FNDs and 0.1 mg $mL^{-1}$ for 120 nm FNDs) acquired using 520 nm laser excitation in a custom-built setup. The oxidized 18 nm FNDs exhibit a typical $NV^0$ PL spectrum, the intensity of which decreases to below the detection threshold upon hydrogenation in this experiment. The oxidized 120 nm FNDs show significant PL from both $NV^0$ and $NV^-$ charge states. Upon hydrogenation, the PL intensity decreases by 39%, but the spectral shape remains unchanged. To rationalize these pronounced differences between the two particle sizes, we employ a simple one-dimensional band bending model to estimate the NV charge state across a 1D cross section of the particles (see SI Figure S4 and File S5 for details). Extrapolating this model to a spherical particle, we find that all NVs inside a particle of 18 nm diameter are sufficiently close to the particle surface to switch from $NV^0$ to $NV^+$ upon hydrogenation as illustrated in Figure 3e) (see SI Figure S6 for details). For the 120 nm FNDs, on the other hand, with an 8 nm depletion layer, only 35% of NVs are within this near-surface volume, in good qualitative agreement with the PL decrease observed in Figure 3d). Interestingly, the fact that the spectral shape of the NV spectrum remains unchanged suggests that the hydrogenation 'switches off' the near-surface NVs while the relative contribution of $NV^0$ and $NV^-$ emission in the 'core' of the particle remains unchanged.

We then tested whether the observed reduction in PL in the 18 nm FND powder is caused by NV charge state switching or annealing-induced damage. 18 nm FND-Hyd particles were drop-cast onto a quartz substrate and imaged before and after a UV-ozone treatment. We find that, on average, the 18 nm FND PL increases 4.0-fold after the UV-ozone treatment. See SI Figure S7 for details. This demonstrates that the decrease in PL observed in Figure 3d) is reversible and can mostly be attributed to a change in the NV charge state and not annealing-induced damage to the diamond lattice.
Previous studies have investigated controlling the NV charge state in nanodiamonds via the particles' surface chemistry[18,32,33], often with the goal of increasing the concentration of the $NV^-$ charge state for spin-based sensing applications. This includes single particle experiments[34,35], and processing performed on particle ensembles and powders[29,36]. The results presented in this and the previous section constitute the first demonstration of reversible switching of NV centers in nanodiamonds between their non-fluorescent $NV^+$ and fluorescent $NV^0$ charge states at the single-particle level (section 2.1) and in FND powders (section 2.2).

**2.3 Widefield voltage sensing and imaging.** Having established that annealing 18 nm FND powders in hydrogen forming gas converts most NV centers into their positive charge state, we investigated the response of the particles' PL to applied voltages. 18 nm FND-Hyd particles were spin-coated onto indium-tin-oxide (ITO) coated cover glass, placed in an aqueous electrochemical cell containing 170 mM NaCl, and imaged with a custom-built wide-field fluorescence microscope (532 nm laser excitation, PL collected above 580 nm) through the glass substrate as illustrated in Figure 4a). An external voltage was applied between the ITO electrode and a platinum counter electrode and the change in FND PL was monitored over time using a CCD camera. See SI Figure S8 for details. We define the change in PL ($\Delta PL$) as

$$\Delta PL = \frac{(I - I_0)}{I_0} \quad (1)$$

where $I$ and $I_0$ are the collected PL intensity in photocounts per second in the presence of an applied voltage and zero volts applied, respectively.

Figure 4b) shows $\Delta PL$ of FND-Hyd (red trace) and FND-Oxy (blue trace) as a function of the applied voltage, which was ramped from +1V to -1V at a rate of 2 mV/s, relative to 0 V applied. For both particle types, $\Delta PL$ of a single pixel at the location of an FND was analyzed (see SI Figures S9 to S11 for PL images). For FND-Hyd and FND-Oxy, we chose the particle with the strongest response in the FOV. The FND-Hyd signal increases by 42% at -1V and decreases monotonically to $\Delta PL$ of –23% at +1V. The FND-Oxy particles, on the other hand, only show a small change of -2% and 1% at -1 and +1 V, respectively. Based on the slope of the FND-Hyd trace between -0.2 and +0.2 V, we determine the FND-Hyd shot-noise limited voltage sensitivity[18] as

$$\eta = \frac{\Delta V}{\Delta PL \sqrt{I_0}} \quad (2)$$

where $\Delta PL$ is the change in PL as defined in equation 1, and $\Delta V$ the change in voltage. This yields a sensitivity of 16 mV $Hz^{-1/2}$ for one of the best-performing FND aggregates.

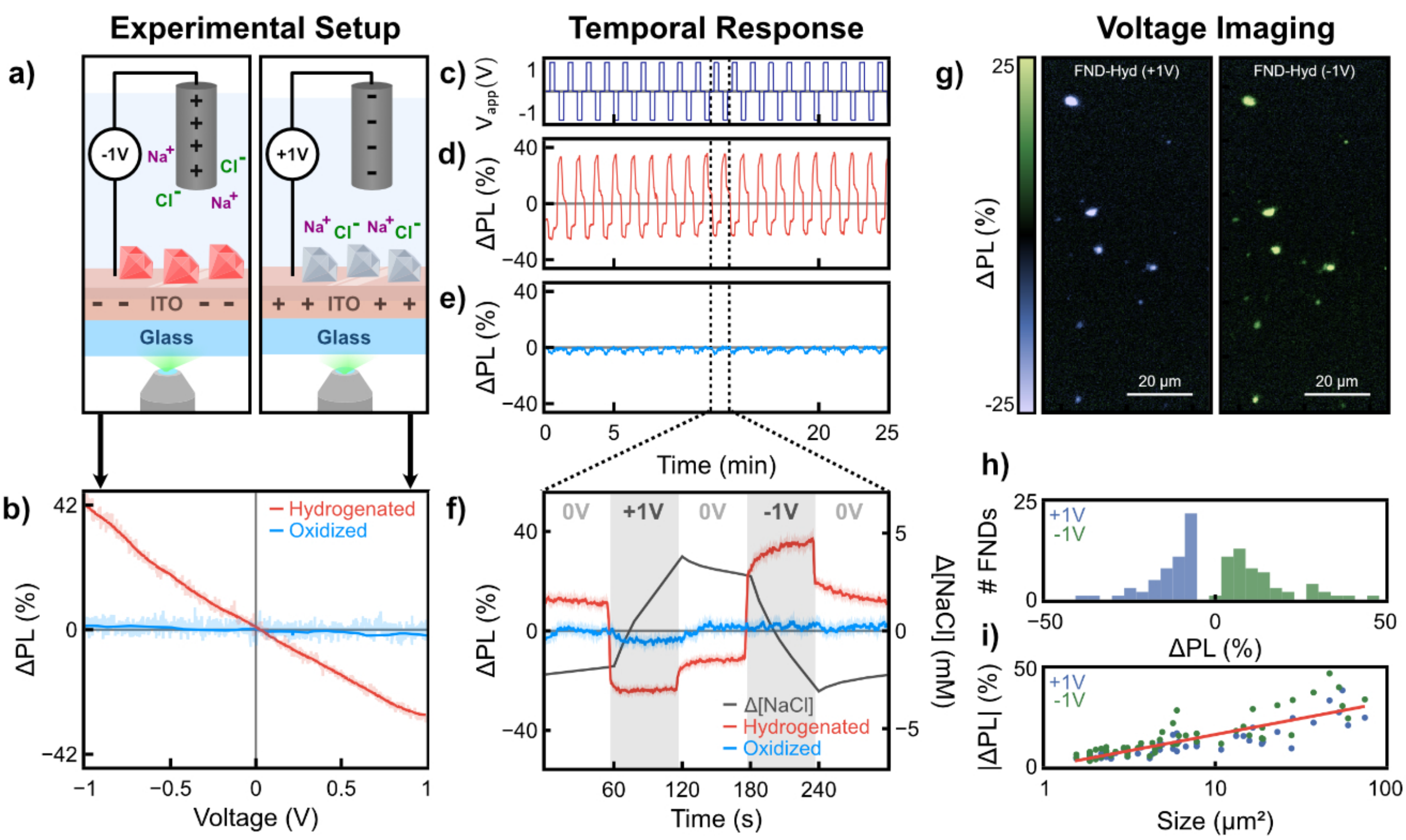

**FIGURE 4** | Voltage response of hydrogenated and oxidized FNDs in an aqueous electrochemical cell. **a)** Graphical representation of the electrochemical cell setup and FNDs dispersed on an ITO-coated glass substrate. **b)** Change in FND-Hyd (red trace) and FND-Oxy (blue trace) PL as a function of applied voltage. **c)** Applied voltage versus time for the experiments shown in d) and e). **d,e)** Change in PL as a function of time for each FND type as voltage pulses are applied. **f)** Change in PL for both FND types and calculated NaCl concentration at the position of FNDs as a function of time for one pulse sequence. **g)** Wide-field PL images of $\Delta PL$ of FND-Hyd particles at the end of a +1 V (left) and -1 V pulse (right). **h)** Histogram of $\Delta PL$ for all particles shown in panel g) for +1V and -1V applied. **i)** Scatter plot of $\Delta PL$ for all particles in g) as a function of the substrate area covered by the FND particle aggregate.

Overall, this change can be explained by a voltage-induced modulation of the NV charge state and aligns qualitatively with previous reports[17,18]. FNDs are in direct contact with the ITO electrode. Applying a negative voltage (Figure 4a, left) pulls positive charges towards the diamond surface, reducing the density of negative acceptors. This, in turn, decreases the total positive charge accumulated in the diamond (reducing the upward band bending), leading to a PL increase caused by $NV^+ \rightarrow NV^0$ conversion. The same -1 V applied voltage has a much smaller effect on FND-Oxy particle PL, suggesting that the additional voltage-induced upward band-bending doesn't significantly affect the charge state equilibrium in the oxidized particles. Applying a positive voltage has the opposite effect; it attracts additional negative charge to the FND surface, resulting in increased buildup of positive charge within the diamond (increased upward band bending), and leads to a decrease in $NV^0$ PL from FND-Hyd particles via $NV^0 \rightarrow NV^+$ conversion. Because $NV^0$ centers in FND-Hyd particles are already depleted, additional downward bending is expected to have a less significant impact, which may explain the asymmetry in the voltage response, such as the lower maximum $|\Delta PL|$ of 29% at +1 V compared to 42% at -1 V.

We then modulated the voltage between 0 V, +1V, and -1V in discrete steps of 20s duration each (Figures 4 c-e) to investigate the reproducibility of the voltage-induced PL change over time and to study the time dynamics. Figures 4d) and 4e) show $\Delta PL$ for FND-Hyd and FND-Oxy, respectively, for 36 voltage pulses over a total acquisition time of 25 minutes. FND-Hyd particles show a highly reproducible change in PL of 36% (-1V) and -22% (+1V) with a standard deviation of 4.2% and 5.3% between individual pulses, respectively. We also find that the maximum PL intensity of the FND-Hyd particles increases by only by 4.0% during the 25-minute measurement, highlighting the exceptional photostability[37] of FNDs. The voltage-induced changes in PL of the FND-Oxy particles remain below 1% (-1 V) and -5% (+1V), while the higher (compared to FND-Hyd) mean PL intensity remains stable.

We then investigated the $\Delta PL$ time dynamics. Since the $\Delta PL$ dynamics in Figures 4d) and 4e) are nearly identical for each pulse, we focused on one +1V and -1V sequence as shown in Figure 4f) and extended the pulse durations to 60 sec each to allow the system to equilibrate for longer. The minimum and maximum $\Delta PL$ values observed for both particle types are consistent with those in Figures 4b), 4d) and 4e). Interestingly, the FND-Hyd ΔPL signal does not return to zero when 0 V is applied after +1 V or -1 V period. Similarly, after a fast initial decrease and increase of $\Delta PL$ within ~100 ms after application of a +1V and -1V voltage pulse, respectively, the PL continues to decrease (8.2% total) or increase (5.4% total), throughout the pulse (see SI Figure S12 for a more detailed analysis).

While the electrical potential at the electrodes is expected to reach equilibrium within a few milliseconds, ion diffusion between electrodes separated by several millimeters, as in our experiment, occurs over seconds. Therefore, we calculated the NaCl concentration as a function of time in three dimensions using a finite-element time-domain model of our electrochemical cell based on the Nernst-Planck equations (see SI Figure S23 for details). The grey trace in Figure 4b) shows the calculated change in NaCl concentration just above the ITO electrode as a function of time. The most significant changes in NaCl, up to 7 mM, happen during the voltage pulses, as expected, and do not reach equilibrium during pulses of either polarity. When 0 V is applied, the NaCl concentration slowly equilibrates toward zero, driven only by the local concentration gradient. Interestingly, the time dynamics of this equilibration during the 0 V periods mirror the dynamics of the FND-Hyd $\Delta PL$ traces, suggesting that ion diffusion may play a more important role than previously thought and will be investigated in Figure 5.

We then explored the feasibility of FND-based voltage imaging using a wide-field microscope. Figure 4g) shows an image of $\Delta PL$ of FND-Hyd particles dispersed on an ITO-coated glass substrate with +1 V and -1 V applied as illustrated in Figure 4a), left. The image is one frame of a video (see SI video File S19) acquired at 10 frames per second over many minutes during which the applied voltage is modulated as shown in Figure 4c). The image shows the change in PL at the end of a +1 V (left) and -1 V pulse (right), relative to 0 V. All 98 particles and particle aggregates in the FOV (see SI Figure S11 for the entire FOV analyzed) show an increase in PL upon application of -1V and a decrease for +1V applied. Figure 4h) shows a histogram of *ΔPL* for all FNDs in the FOV for +1V and -1V applied. While both polarities show a broad range of *ΔPL* values, the distributions indicate that all particles exhibit a reliable, consistent response, with the average particle showing *ΔPL* values of -11% and 15% upon application of +1 V and -1 V, respectively. We also find that the magnitude of *ΔPL* correlates with the FND aggregate size: larger aggregates typically exhibit a larger *ΔPL* (Figure 4i). Overall, the magnitude of the change in PL observed here is in good agreement with that reported by Karaveli et al for hydrogenated FNDs of comparable size[17]. However, while particles in the previous study showed a range of different responses, e.g. many particles responded only to positive or negative voltages, FND-Hyd investigated here all show the same response, which is important for applications.

**2.4 Widefield Ion Concentration Imaging**. To further study the effect of changes in the local ion concentration, we investigated an electrode geometry where FNDs are not in direct contact with an electrode, as illustrated in Figure 5a). Platinum electrodes were deposited on quartz substrates using photolithography and physical vapor deposition and FND-Hyd particles were self-assembled between the electrodes. The substrates were placed in an aqueous electrochemical cell containing 1.7 M NaCl, voltages of 0, +1, and -1 V were applied, and changes in FND PL were monitored using a wide-field fluorescence microscope. See SI Figures S14 & S15 for details.

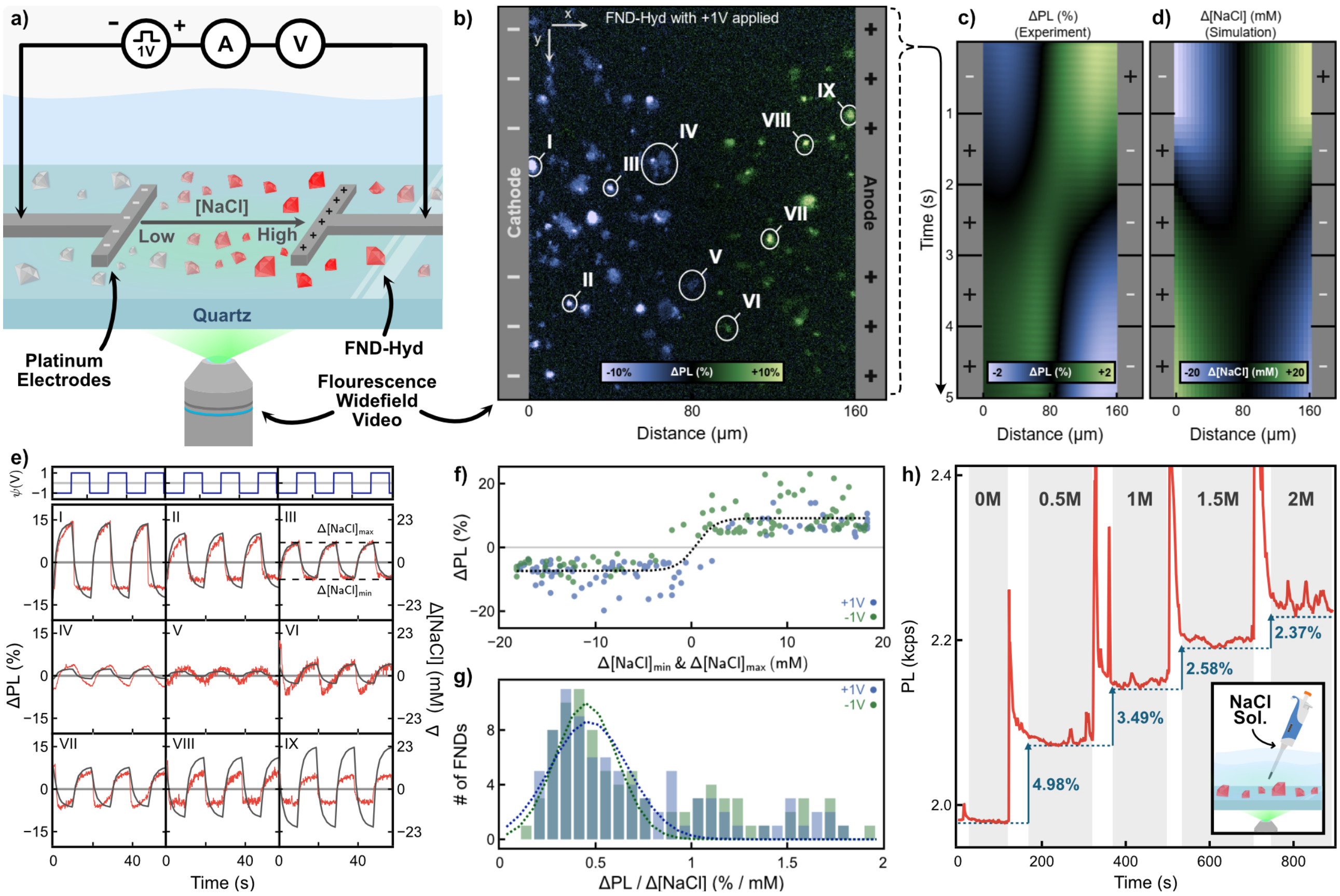

**FIGURE 5 |** Modulation of the PL of hydrogenated nanodiamonds by voltage-induced microscale salt concentration gradients. **a)** Schematic illustration of the experimental setup. **b)** Wide-field PL image of the change in PL of FND-Hyd particles dispersed on a quartz substrate between two platinum electrodes with +1 V applied. **c)** ΔPL binned along the direction of the electrodes in b) as a function of time as the polarity of the applied voltage changes from +1 V (0 - 1 s) to -1V (1-5 s). **d)** Calculated change in NaCl concentration (Δ[NaCl]) between the electrodes as a function of time for the applied voltages and duration shown in c). Δ[NaCl] was calculated using FDTD simulations. **e)** ΔPL (left axes) and calculated Δ[NaCl] (right axes) a as a function of time for locations I to IX indicated in panel b) as the polarity of the applied voltage changes. **f)** ΔPL of all FND-Hyd particles between the electrodes as a

function of the minimum and maximum calculated change in NaCl concentration (Δ[NaCl]max and Δ[NaCl]min, respectively) during +1V (purple dots) and -1V pulses (green dots). **g)** Histogram of ΔPL normalized by the calculated Δ[NaCl] for all particles. **h)** PL intensity averaged over 156 x 156 μm2 of a quartz substrate covered with FND-Hyd particles versus time as the salt concentration is increased stepwise via the addition of a concentrated NaCl solution.

Figure 5b) shows an image of the change in FND-Hyd PL (*ΔPL*) after +1 V was applied for 10 s relative to 0 V applied. The two electrodes were separated by 160 μm and are illustrated on the left and right sides of the image for clarity. FNDs and FND aggregates are distributed across the field of view between the electrodes. FNDs on the left and right halves of the image show a pronounced decrease of up to -11% and an increase of up to 17% in PL, respectively. Qualitatively, these changes coincide with an NaCl concentration gradient from left (lower NaCl) to right (higher NaCl) created by the applied voltage. Given the high salt concentration of 1.7 M and the resulting Debye length of ~0.23 nm, strong ionic screening will occur at the electrode surfaces, and only a small and constant bulk electric field is expected at the locations of all investigated FNDs, which will be discussed in more detail below.

To quantitatively investigate the relationship between changes in PL and NaCl concentration, we calculated the NaCl concentration (*[NaCl]*) for this electrode geometry (see previous section) and directly compared *ΔPL* and the calculated change in NaCl concentration (*Δ[NaCl]*)as a function of time as the electrode polarity was reversed. Figure 5c) shows FND-Hyd ΔPL for all x-positions between the electrodes, binned vertically along the electrode y-axis in Figure 5b), over time as the applied potential is reversed at *t* = 1 s. A Gaussian blur filter was applied ($\sigma_{\text{distance}} = 15\ \mu\text{m}, \sigma_{\text{time}} = 1.36\ \text{s}$) to clearly visualize spatiotemporal dynamics and enable direct comparison with simulations. Figure 5d) displays *Δ[NaCl]*) at each x-position between the electrodes as a function of time. Before the voltage reversal at *t* = 1 s, both FND-Hyd PL and salt concentration decrease near the cathode (left) and increase near the anode (right), consistent with the conditions shown in Figure 5b). The amplitude of the *ΔPL* signal in Figure 5c) is reduced by the Gaussian filtering. After the polarity reversal, both *ΔPL* and *Δ[NaCl]* drop to zero within approximately 1.5 s near both electrodes, then change sign, increasing at the anode (left) and decreasing at the cathode (right), approaching values similar to those at *t* = 0 s but with opposite signs. Overall, Figures 5c) and 5d) show that the spatiotemporal dynamics of the change in FND-Hyd PL closely follow that of the calculated NaCl concentration.

To investigate this similarity more quantitatively and locally, we determined *ΔPL* as a function of time over three ±1 V voltage cycles for the nine locations indicated in Figure 5b). Figure 5e) displays this change in PL versus time (red traces, left axes) and the change in NaCl concentration (grey trace, right axes) calculated at these locations. For many locations at different x-positions between the electrodes, e.g., I, III, VI, and VII, the time dynamics and amplitudes of the PL increase closely follow changes in NaCl concentration. Overall, the negative changes in PL at these locations follow *Δ[NaCl]* but exhibit slightly different dynamics; *ΔPL* decreases more rapidly and saturates more quickly than *Δ[NaCl]*. In some locations, e.g., VIII and IX, the dynamics partially match as described above, but the amplitudes of *ΔPL* are smaller for a given *Δ[NaCl]*.

We then investigate in greater detail the relationship between the amplitude of *ΔPL* and the local *Δ[NaCl]*. Figure 5f) shows the maximum and minimum *ΔPL* ($\Delta PL_{min}$ and $\Delta PL_{max}$, see Figure 5e) for all FNDs and FND aggregates with $\Delta PL_{max}$ - $\Delta PL_{min}$ above 4% as a function of the calculated *Δ[NaCl]* in that location for +1 V (blue dots) and -1 V (green dots) applied. While there is a significant variation in ΔPL amplitudes for all *Δ[NaCl]* values, it is important to note that all FNDs respond to positive and negative changes in salt concentration with an increase and decrease in PL, respectively. On average, we observe a continuous, nonlinear relationship between *ΔPL* and *Δ[NaCl]*, illustrated by a sigmoidal fit to the experimental data in Figure 5f) (dashed black line). Figure 5g) shows a histogram for $\Delta PL_{min}$ and $\Delta PL_{max}$ divided by $\Delta[NaCl]$ for all spots plotted in Figure 5f). Distributions for +1V and -1V applied peak at 0.45 % $mM^{-1}$ and 0.31 % $mM^{-1}$, respectively, while a few FNDs show a response of up to 1.5 % $mM^{-1}$ and above.

All results presented so far suggest that changes in NaCl concentration modulate the PL intensity of FND-Hyd particles. However, the NaCl gradients investigated in Figures 5a) - 5g) were created by applying an external voltage, which produces a very weak and constant electric field between the electrodes screened by NaCl and results in a small ionic current ($J$). We measured $J$ = 38 nA, suggesting a residual electric field of 2 V/m in our electrode geometry (see SI Figure S15). The electric field is small, and based on experiments we reported for a solid-state system, unlikely to cause a measurable change in PL[28]. The electric field also has a constant magnitude between the electrodes and hence cannot explain the PL gradients observed here. While the potential effect of the ionic current is even less understood, the fact that it is also constant suggests that it alone cannot explain the observed PL changes, leaving the salt concentration as the most likely cause.

Hence, we monitored the PL of FND-Hyd particles as a function of NaCl concentration in bulk water in real time without applying an external voltage, as shown in Figure 5h). The average PL intensity over a 156 x 156 µm region of the FND-coated substrate was measured as a concentrated salt solution was added incrementally, and the NaCl concentration in solution varied between 0 and 2 M. Upon addition of NaCl, the PL intensity temporarily spiked due to mechanical motion of the system and then settled within a few seconds. After each NaCl addition, we observe a clear increase in FND PL of up to 4.9% after the first addition, which decreases to 2.3% after the last addition.

The experiment shown in Figure 5h) clearly demonstrates that changes in salt concentration alone modulate the NV PL intensity of FND-Hyd particles. The sign of this change, i.e. an increase in salt leads to an increase in PL, is consistent with all experiments shown in Figure 5. However, while we observe an average response of ~0.4 % $mM^{-1}$ in the experimental geometry illustrated in Figure 5a), we observe a much lower response below 0.01 % $mM^{-1}$ in the experiment in Figure 5h) in a similar salt concentration range (1 M to 1.5 M). Understanding the discrepancy between these two experiments and, more fundamentally, the physical mechanisms underlying the salt-induced modulation of NV PL will be the focus of future studies and an important step towards developing FNDs as sensitive all-optical ion sensors. Based on our results, we hypothesize that ion-induced changes in the electric double layer at the diamond-electrolyte liquid interface play an important role in the observed modulation of the NV charge state.

**3. Conclusion**. We have demonstrated that the NV charge state in sub-30 nm FNDs can be switched from $NV^0$ in oxidized FNDs to $NV^+$ via surface hydrogenation. The hydrogenation was shown to be effective for small individual FND aggregates dispersed on a solid substrate (Figure 2) and for FND powder at the 100 mg scale (Figure 3). In both cases, a simple UV-ozone treatment re-oxidized the FND surface, leading to a recovery of the FND PL. In experiments in electrochemical cells, we showed that the residual $NV^0$ PL of hydrogenated FNDs is strongly modulated by applied voltages and by the concentration of ions in solution. The voltage dependence was investigated for hydrogenated FNDs in direct contact with a transparent conductive electrode immersed in an electrolyte solution, with a voltage applied via a counter electrode in the electrolyte. For all investigated particles, we observed strong increases and decreases in PL upon application of negative and positive voltages, respectively, with a sensitivity of up to 16 mV $Hz^{-1/2}$ (Figure 4). These PL changes can be explained by a voltage-induced switching of the NV charge state between $NV^0$ and $NV^+$. Lastly, we examined the effect of solution salt concentration on the PL of FND-Hyd in a horizontal electrode geometry, where the FNDs are not in contact with the electrodes, which create an NaCl gradient upon application of a voltage (Figure 5). We discovered that the PL intensity is modulated by changes in NaCl concentration with a sensitivity of up to 1.8% $\Delta PL$ per mM NaCl. Our study provides critical new insights into the control and properties of NV centers in sub-30 nm FNDs, thereby paving the way for their technological development as scalable, cost-effective and all-optical nanoscale voltage and ion concentration sensors.

## Experimental Section

**Nanodiamond Materials and Processing**. For experiments shown in Figure 2, commercially available oxidized 20 nm FNDs containing ~1 ppm NV centers (Adamas Nanotechnologies, USA) were used as received (1 mg $mL^{-1}$ aqueous suspension) and spin-coated at 2000 rpm for 1 min onto a marked quartz substrate (see below). For experiments shown in Figures 3 to 5, FND powders with 18 nm (Pureon, Switzerland) and 120 nm (Nabond, China) nominal particle size were irradiated with 2 MeV electrons to a fluence of $1 \times 10^{18}$ $cm^{-2}$, annealed in argon at 900°C for 2h, and annealed in air at 520°C for 2.5h to create NV centers and an oxidized surface (FND-Oxy). 150 mg of each powder was annealed in a tube furnace in high-purity forming gas (5% $H_2$, 95% $N_2$) at 800°C for 1 hour for hydrogenation (FND-Hyd). All particles were dispersed in deionized (DI) water (1-10 mg $mL^{-1}$) using probe sonication (120 W, 1h, 66% duty cycle). The FND powders and suspensions described here were used in experiments shown in Figure 3.

**Sample Fabrication.** For the FND-Hyd self-assembly discussed in Figure 1, a quartz substrate was cleaned via rinsing in acetone, ethanol and water, followed by UV-ozone cleaning (10 min), and immersed in an FND-Hyd suspension (10 mg $mL^{-1}$, 10 min), rinsed with DI water and dried under nitrogen flow.

For experiments shown in Figure 2, platinum registration markers were fabricated on a quartz substrate using photolithography and electron beam physical vapor deposition (see SI Figure S12) and 20 nm oxidized FNDs (1 mg $mL^{-1}$, Adamas Nanotechnologies, USA) were attached electrostatically using self-assembled by submerging the substrates in 10 mg $mL^{-1}$ polyallylamine hydrochloride (PAH) solution and 1 mg $mL^{-1}$ FND-Oxy solution both for 10 minutes, then annealing at 400 °C in a single-zone tube furnace (MTI, United States) in air for 1 hour to remove residual PAH. The marked substrate with FNDs was then hydrogenated in a tube furnace in forming gas (5% $H_2$, 95% $N_2$, 800°C, 1 h) to create FND-Hyd, then exposed to ozone for 10 minutes in a UV-ozone cleaner to create FND-Ozo particles.

For experiments in Figure 4, an indium-tin-oxide (ITO) coated glass coverslip (0.17 mm, SPI Supplies, USA) was cleaned (rinsing in acetone, ethanol and water, followed by UV-ozone cleaning for 10 min). Suspensions of 18 nm FND-Oxy and FND-Hyd (both 1 mg $mL^{-1}$) were spin coated onto the substrate (1000 rpm for 10 s then 4000 rpm for 30 s) and substrates dried on a hot plate (110 °C for 10 min).

For experiments in Figure 5, platinum electrodes 500 µm wide and 160 µm apart (see Figure S11) were fabricated on a quartz substrate as described above. The truncated top section of a centrifuge tube was glued to the substrate using silicone adhesive to create a watertight fluid well surrounding the electrodes. FND-Hyd particles were self-assembled on the substrate by filling the well with FND-Hyd particle suspension (10 mg $mL^{-1}$), particles allowed to attach to the substrate surface for 10 min, then rinsed with DI water, and dried in air.

**Materials and Characterization**. AFM images were acquired using an Asylum MFP-3D Infinity Atomic Force Microscope (Oxford Instruments, UK), using a Tap-300AL-G AFM tip (Budget Sensors, Bulgaria) in tapping mode. FTIR spectra were acquired using a Frontier NIR spectrometer (Perkin Elmer, USA) with a diamond crystal attenuated total reflectance (ATR) attachment. Dynamic light scattering and zeta potential measurements were performed with a Zeta Sizer ZS Nano (Malvern Panalytical, UK).

**Confocal PL imaging and Spectroscopy.** Confocal PL images and spectra were acquired using a custom-built confocal microscope (see SI Figure S2 for schematics) using a tunable white light laser (Fianium, NKT Photonics, Denmark) at 532 nm / 10 nm bandwidth, with a 532 nm short-pass filter for excitation through a 100 × air objective (TU PlanApo, NA=0.90, Nikon, Japan) and collected through a 550 nm long-pass filter, fiber-coupled, then split 25%/75% into an avalanche photodiode (Excelitas, USA) and a spectrometer (Princeton Instruments, USA). The objective was raster scanned across the sample for imaging and positioned to investigate individual particles using a nano-positioner (Physik Instrumente, Germany).

**Electrochemical Cell Experiments.** For experiments shown in Figure 4, an FND-coated ITO-substrate (see section 4.2) was submerged in 0.17 M NaCl solution in a custom-built electrochemical cell. Voltages were applied using a signal generator (Rigol, China) via a platinum wire in contact with the ITO surface. A 25 mm-wide platinum sheet submerged in the electrolyte was used as the counter-electrode. The electric current through the electrochemical cell was converted to a voltage using a pico-ammeter (Keithley, Tektronix, USA). The applied voltage and image acquisition (see below) were controlled and synchronized via a DAQ (National Instruments, USA) and a LabView.

Wide-field PL imaging was performed using a custom-built inverted fluorescence microscope based on a commercial microscope frame (Eclipse, Nikon, Japan). A collimated laser beam (532 nm, ~400 mW, continuous wave) was focused on the back aperture of a 40 × objective (NA 1.4, Nikon, Japan) with adjustable NA (to account for different substrate thicknesses) for wide-field illumination of samples. PL was collected through the same objective, separated from the excitation beam using a dichroic (532 nm) and long-pass (580 nm) optical filters, and detected with a camera (Andor Neo, Oxford Instruments, UK). For the experiments shown in Figure 5, the system described in the previous two paragraphs was used to investigate voltage -induced salt gradients. To measure the change in PL as a function of ionic strength only (no voltages applied), the NaCl concentration was increased stepwise from 0 M to 2 M by adding specific volumes of concentrated NaCl solution (3 M), such that concentration increases in steps of 0.5M. The salt concentration was allowed to equilibrate diffusively for 200s between additions.

**Numerical Simulations**. 2D time-dependent simulations of the ion concentration in the voltage and ion-concentration sensing electrochemical cell experiments were performed using the 'Tertiary Current Distribution (Nernst Planck)' interface for COMSOL Multiphysics 6.2 (See SI Section S23 for details).

**Acknowledgements.** PR acknowledges support through an Australian Research Council (ARC) DECRA Fellowship (DE200100279), ARC Discovery Projects (DP220102518, DP250100125), and an RMIT University Vice-Chancellor's Senior Research Fellowship. This work was supported with funding from the US Air Force Office of Scientific Research (FA9550-25-1-0260). This work was performed in part at the RMIT Micro Nano Research Facility in the Victorian node of the Australian National Fabrication Facility (ANFF) and the RMIT Microscopy and Microanalysis Facility (RMMF).

**References**

[1] S. Sotoma, C. P. Epperla, H.-C. Chang, *ChemNanoMat* **2018**, *4*, 15.

[2] P. Neumann, I. Jakobi, F. Dolde, C. Burk, R. Reuter, G. Waldherr, J. Honert, T. Wolf, A. Brunner, J. H. Shim, D. Suter, H. Sumiya, J. Isoya, J. Wrachtrup, *Nano Lett.* **2013**, *13*, 2738.

[3] D. A. Simpson, E. Morrisroe, J. M. McCoey, A. H. Lombard, D. C. Mendis, F. Treussart, L. T. Hall, S. Petrou, L. C. L. Hollenberg, *ACS Nano* **2017**, *11*, 12077.

[4] Y. Wu, M. N. A. Alam, P. Balasubramanian, A. Ermakova, S. Fischer, H. Barth, M. Wagner, M. Raabe, F. Jelezko, T. Weil, *Nano Lett.* **2021**, *21*, 3780.

[5] T. F. Segawa, R. Igarashi, *Prog. Nucl. Magn. Reson. Spectrosc.* **2023**, *134–135*, 20.

[6] J.-P. Tetienne, T. Hingant, L. Rondin, A. Cavaillès, L. Mayer, G. Dantelle, T. Gacoin, J. Wrachtrup, J.-F. Roch, V. Jacques, *Phys. Rev. B* **2013**, *87*, 235436.

[7] A. Ermakova, G. Pramanik, J.-M. Cai, G. Algara-Siller, U. Kaiser, T. Weil, Y.-K. Tzeng, H. C. Chang, L. P. McGuinness, M. B. Plenio, B. Naydenov, F. Jelezko, *Nano Lett.* **2013**, *13*, 3305.

[8] "Detection of atomic spin labels in a lipid bilayer using a single-spin nanodiamond probe," DOI 10.1073/pnas.1300640110can be found under https://www.pnas.org/doi/10.1073/pnas.1300640110, **n.d.**

[9] T. Fujisaku, R. Tanabe, S. Onoda, R. Kubota, T. F. Segawa, F. T.-K. So, T. Ohshima, I. Hamachi, M. Shirakawa, R. Igarashi, *ACS Nano* **2019**, *13*, 11726.

[10] T. Rendler, J. Neburkova, O. Zemek, J. Kotek, A. Zappe, Z. Chu, P. Cigler, J. Wrachtrup, *Nat. Commun.* **2017**, *8*, 14701.
[11] N. Aslam, H. Zhou, E. K. Urbach, M. J. Turner, R. L. Walsworth, M. D. Lukin, H. Park, *Nat. Rev. Phys.* **2023**, *5*, 157.
[12] T. Zhang, G. Pramanik, K. Zhang, M. Gulka, L. Wang, J. Jing, F. Xu, Z. Li, Q. Wei, P. Cigler, Z. Chu, *ACS Sens.* **2021**, *6*, 2077.
[13] Y. Wu, T. Weil, *Adv. Sci.* **2022**, *9*, 2200059.
[14] F. Dolde, H. Fedder, M. W. Doherty, T. Nöbauer, F. Rempp, G. Balasubramanian, T. Wolf, F. Reinhard, L. C. L. Hollenberg, F. Jelezko, J. Wrachtrup, *Nat. Phys.* **2011**, *7*, 459.
[15] K. Bian, W. Zheng, X. Zeng, X. Chen, R. Stöhr, A. Denisenko, S. Yang, J. Wrachtrup, Y. Jiang, *Nat. Commun.* **2021**, *12*, 2457.
[16] K. Bian, W. Zheng, X. Zeng, X. Chen, R. Stöhr, A. Denisenko, S. Yang, J. Wrachtrup, Y. Jiang, *Nat. Commun.* **2021**, *12*, 2457.
[17] S. Karaveli, O. Gaathon, A. Wolcott, R. Sakakibara, O. A. Shemesh, D. S. Peterka, E. S. Boyden, J. S. Owen, R. Yuste, D. Englund, *Proc. Natl. Acad. Sci.* **2016**, *113*, 3938.
[18] D. J. McCloskey, N. Dontschuk, A. Stacey, C. Pattinson, A. Nadarajah, L. T. Hall, L. C. L. Hollenberg, S. Prawer, D. A. Simpson, *Nat. Photonics* **2022**, *16*, 730.
[19] B. Grotz, M. V. Hauf, M. Dankerl, B. Naydenov, S. Pezzagna, J. Meijer, F. Jelezko, J. Wrachtrup, M. Stutzmann, F. Reinhard, J. A. Garrido, *Nat. Commun.* **2012**, *3*, 729.
[20] A. Härtl, J. A. Garrido, S. Nowy, R. Zimmermann, C. Werner, D. Horinek, R. Netz, M. Stutzmann, *J. Am. Chem. Soc.* **2007**, *129*, 1287.
[21] C. Schreyvogel, V. Polyakov, R. Wunderlich, J. Meijer, C. E. Nebel, *Sci. Rep.* **2015**, *5*, 12160.
[22] D. Takeuchi, S.-G. Ri, H. Kato, C. E. Nebel, S. Yamasaki, *Phys. Status Solidi A* **2005**, *202*, 2098.
[23] J. A. Garrido, S. Nowy, A. Härtl, M. Stutzmann, *Langmuir* **2008**, *24*, 3897.
[24] C. E. Nebel, B. Rezek, D. Shin, H. Watanabe, *Phys. Status Solidi A* **2006**, *203*, 3273.
[25] V. Petrakova, V. Benson, M. Buncek, A. Fiserova, M. Ledvina, J. Stursa, P. Cigler, M. Nesladek, *Nanoscale* **2016**, *8*, 12002.
[26] M. Sow, H. Steuer, S. Adekanye, L. Ginés, S. Mandal, B. Gilboa, O. A. Williams, J. M. Smith, A. N. Kapanidis, *Nanoscale* **2020**, *12*, 21821.
[27] S. Menon, A. Tyler, M. Mather, *Acad. Quantum* **2025**, *2*.
[28] R. Styles, M. Han, T. Goris, J. Partridge, B. Johnson, B. del Rosal, A. Abraham, H. Ebendorff-Heidepriem, B. Gibson, N. Dontschuk, J.-P. Tetienne, P. Reineck, *Adv. Funct. Mater.* **2025**, *35*, DOI 10.1002/adfm.202512068.
[29] D. J. McCloskey, D. Roberts, L. V. H. Rodgers, Y. Barsukov, I. D. Kaganovich, D. A. Simpson, N. P. de Leon, A. Stacey, N. Dontschuk, *Adv. Mater. Interfaces* **2024**, *11*, 2400242.
[30] K. Chea, E. S. Grant, K. J. Rietwyk, H. Abe, T. Ohshima, D. A. Broadway, J.-P. Tetienne, G. Bryant, P. Reineck, *Adv. Mater. Interfaces* **2026**, e00957.
[31] O. A. Williams, J. Hees, C. Dieker, W. Jäger, L. Kirste, C. E. Nebel, *ACS Nano* **2010**, *4*, 4824.
[32] A. Sadžak, I. Pérez, M. Guimarães, L. Spantzel, F. Jelezko, M. Börsch, A. Krueger, *Adv. Funct. Mater.* **n.d.**, *n/a*, e21387.
[33] L. V. H. Rodgers, S. T. Nguyen, J. H. Cox, K. Zervas, Z. Yuan, S. Sangtawesin, A. Stacey, C. Jaye, C. Weiland, A. Pershin, A. Gali, L. Thomsen, S. A. Meynell, L. B. Hughes, A. C. B. Jayich, X. Gui, R. J. Cava, R. R. Knowles, N. P. de Leon, *Proc. Natl. Acad. Sci.* **2024**, *121*, e2316032121.
[34] L. Rondin, G. Dantelle, A. Slablab, F. Grosshans, F. Treussart, P. Bergonzo, S. Perruchas, T. Gacoin, M. Chaigneau, H.-C. Chang, V. Jacques, J.-F. Roch, *Phys. Rev. B* **2010**, *82*, 115449.
[35] T. W. Shanley, A. A. Martin, I. Aharonovich, M. Toth, *Appl. Phys. Lett.* **2014**, *105*, 063103.
[36] E. Mayerhoefer, A. Krueger, *Acc. Chem. Res.* **2022**, *55*, 3594.
[37] F. Treussart, V. Jacques, E. Wu, T. Gacoin, P. Grangier, J.-F. Roch, *Phys. B Condens. Matter* **2006**, *376–377*, 926.

*Supporting Information for*

# Ion Concentration and Voltage Imaging with Fluorescent Nanodiamonds

Patrick Voorhoeve[1], Hiroshi Abe[2], Takeshi Ohshima[2,3], Qiang Sun[4], Anita Quigley[1,5,6], Rob Kapsa[1,5,6], Nikolai Dontschuk[7], Philipp Reineck[4*]

[1] Biomedical Engineering, School of Engineering, RMIT University, Melbourne, VIC, Australia.

[2] National Institutes for Quantum Science and Technology (QST), Takasaki, Gunma, 370-1292, Japan.

[3] Department of Materials Science, Tohoku University, Aoba, Sendai, Miyagi 980-8579, Japan

[4] School of Science, RMIT University, Melbourne, VIC 3001, Australia

[5] Aikenhead Centre for Medical Discovery, 27 Victoria Parade, Fitzroy, Melbourne, VIC 3065, Australia.

[6] Clinical Neurosciences, St Vincent's Hospital Melbourne, Fitzroy, Melbourne, VIC 3065, Australia

[7] School of Physics, University of Melbourne, Parkville, Victoria 3010, Australia.

*email: philipp.reineck@rmit.edu.au

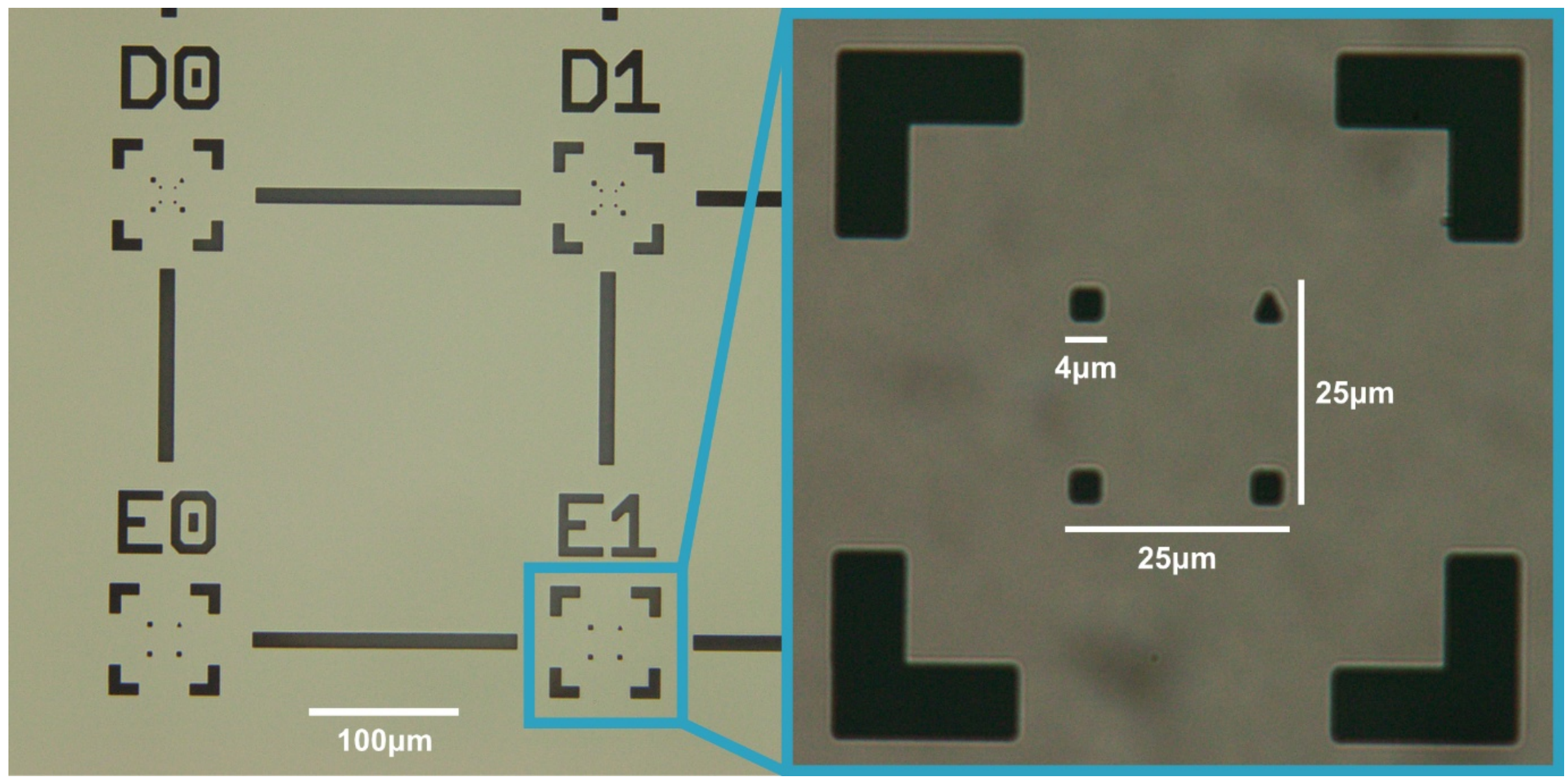

**FIGURE S1** | **(left)** Bright field microscopy images of custom-made platinum registration markers used for experiments in section 2 of the main text. A 1 × 25 × 25 mm fused silica (quartz) substrate was patterned using an MLA-150 maskless aligner (Heidelberg Instruments, Germany), and a 100 nm platinum layer was deposited using an electron-beam physical vapor deposition chamber (Kurt J. Lesker, Germany) then diced into 5 mm squares using a DAD321 dicing saw (DISCO Corporation, Japan) and photoresist removal via sonication in acetone. Final substrate contained many variations of markers, E1 was used for all experiments. **(right)** Expanded bright field microscopy image of E1 region of interest. Markers within this region are 4 µm wide, spaced in a 25 µm grid pattern. Top-right marker is shaped as a triangle to enable identification of orientation.

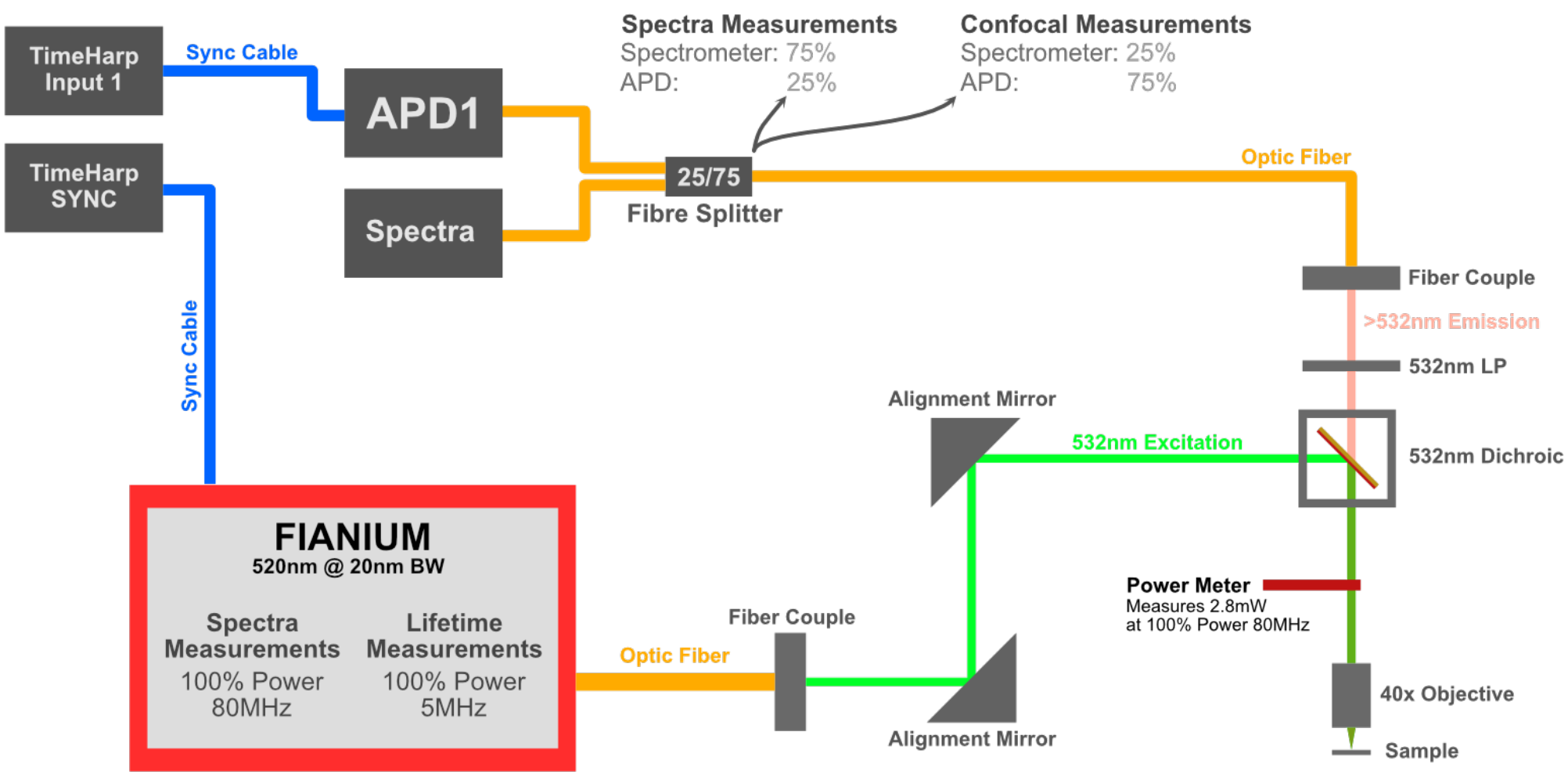

**FIGURE S2** | Diagram of the custom-built confocal microscope used in section 2.1 of main text. Microscope is using a tunable white light laser (Fianium, NKT Photonics, Denmark) at 520 nm / 10 nm bandwidth, with a 532 nm short-pass filter for excitation through a 100 × objective (TUPlanApo, NA=0.90, Nikon, Japan) and collected through a 550 nm long-pass filter, fiber-coupled, then split 25%/75% into an avalanche photodiode (Excelitas, USA) and a spectrometer (Princeton Instruments, USA). The objective was raster scanned across the sample for imaging and positioned to investigate individual particles using a nano-positioner (Physik Instrumente, Germany).

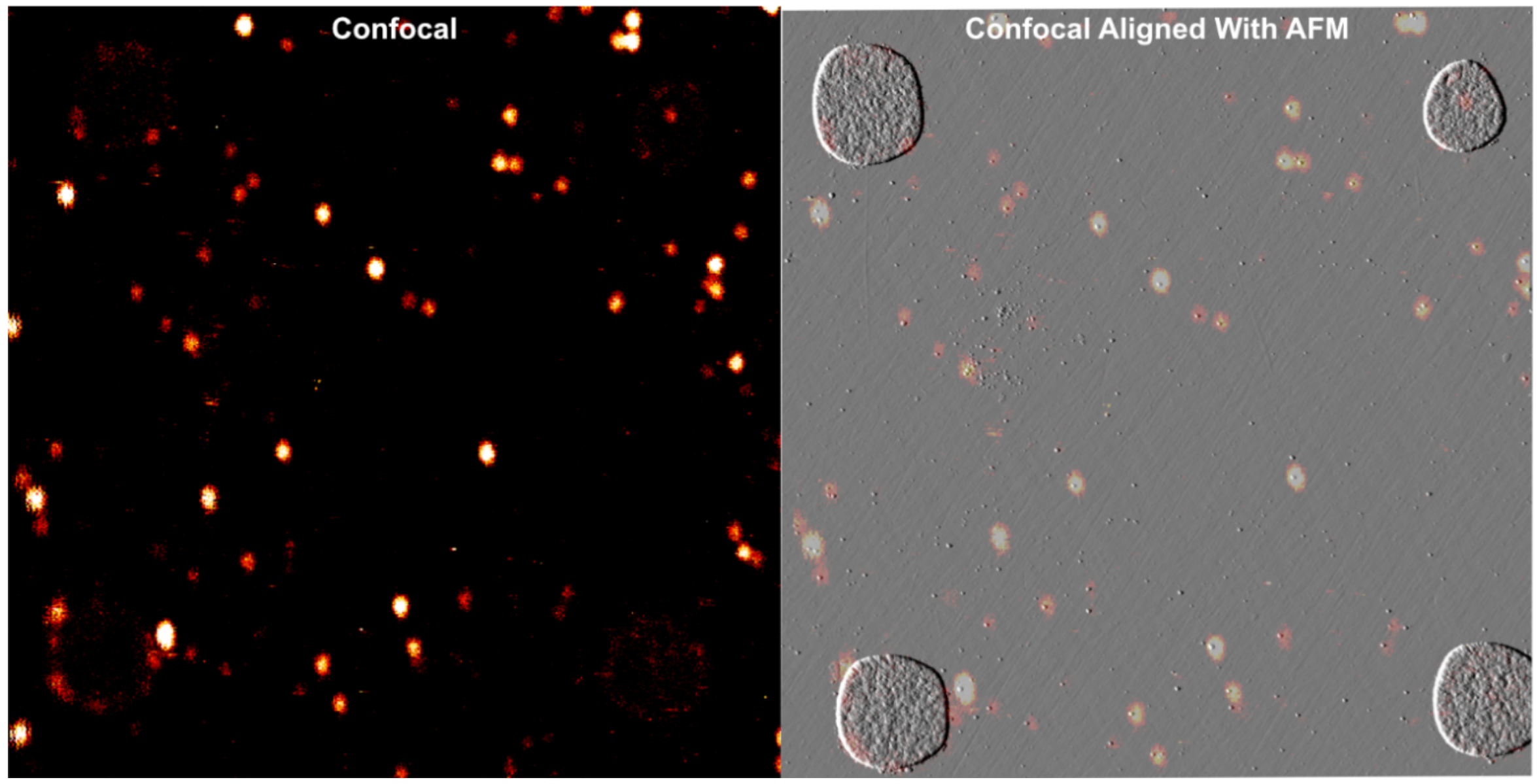

**FIGURE S3 | (left)** Confocal fluorescence image of FND-Hyd particles self-assembled on quartz with four 4 μm platinum registration markers enclosing a 25 μm region of interest. **(right)** AFM amplitude map of the same region of interest as (left) with (left) overlaid. Images were aligned using the registration markers and OpenCV's 2D homography algorithms (https://opencv.org/), which align two images by selecting matching pixels on both images and sequentially applying transformations to minimize the distance between all matching pixel pairs[1,2]. This allows confocal and AFM images to be aligned and directly compared between the particles oxidised, hydrogenated, and UV-Ozone treated stages.

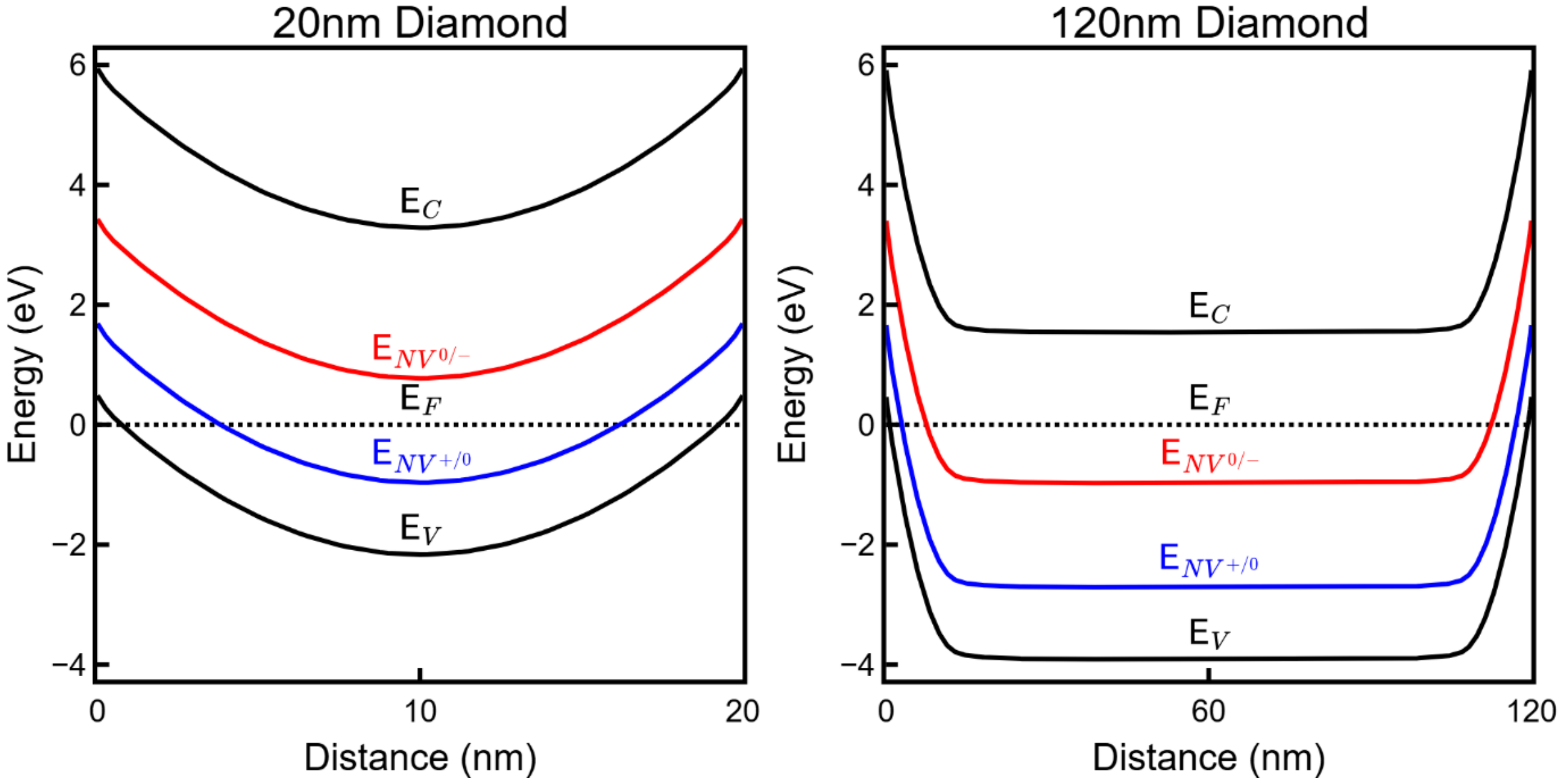

**FIGURE S4** | Results of a semi-quantitative simulated semiconductor model of band-bending within the diamond lattice comparing 20 and 120 nm oxidized and hydrogenated FNDs in a liquid electrolyte. The simulation was created using the semiconductor interface in COMSOL Multiphysics. The simulation domain was a 1D diamond with a 20 nm (left) and 120 nm (right) width, with both simulations using the same parameter sets.

NV centres were modelled as $NV^0$ defects that can accept ($NV^-$) and donate electrons ($NV^+$) using 'Trap-Assisted-Recombination nodes' and discrete NV charge state transition energies reported in literature for diamond[3]. Substitutional nitrogen ($N_S^0$) was modelled as an electron donor. We assumed NV and $N_S^0$ concentrations of 1 ppm and 80 ppm, respectively. The charge of surface adsorbates was modelled by applying a 'Surface Charge Density' boundary condition and set to a value of $\sigma = -0.05\ \mathrm{C.m^{-2}}$. The exact physical value of the surface charge density is unknown and difficult to estimate. However, sweeping the surface charge density shows the expected 2D-hole-gas region ($E_V > E_F$) when $\sigma > -0.02\ C.m^{-2}$. Thus, we assume that the adsorbate charge density is at this order of magnitude.

**FILE S5** | COMSOL Multiphysics' automatically generated report detailing the simulation setup for results shown in SI Figure S4 is available here:
https://drive.google.com/drive/folders/1PYIRaSqKtedlfrCbFal94lqe9sjNX6Va?usp=sharing

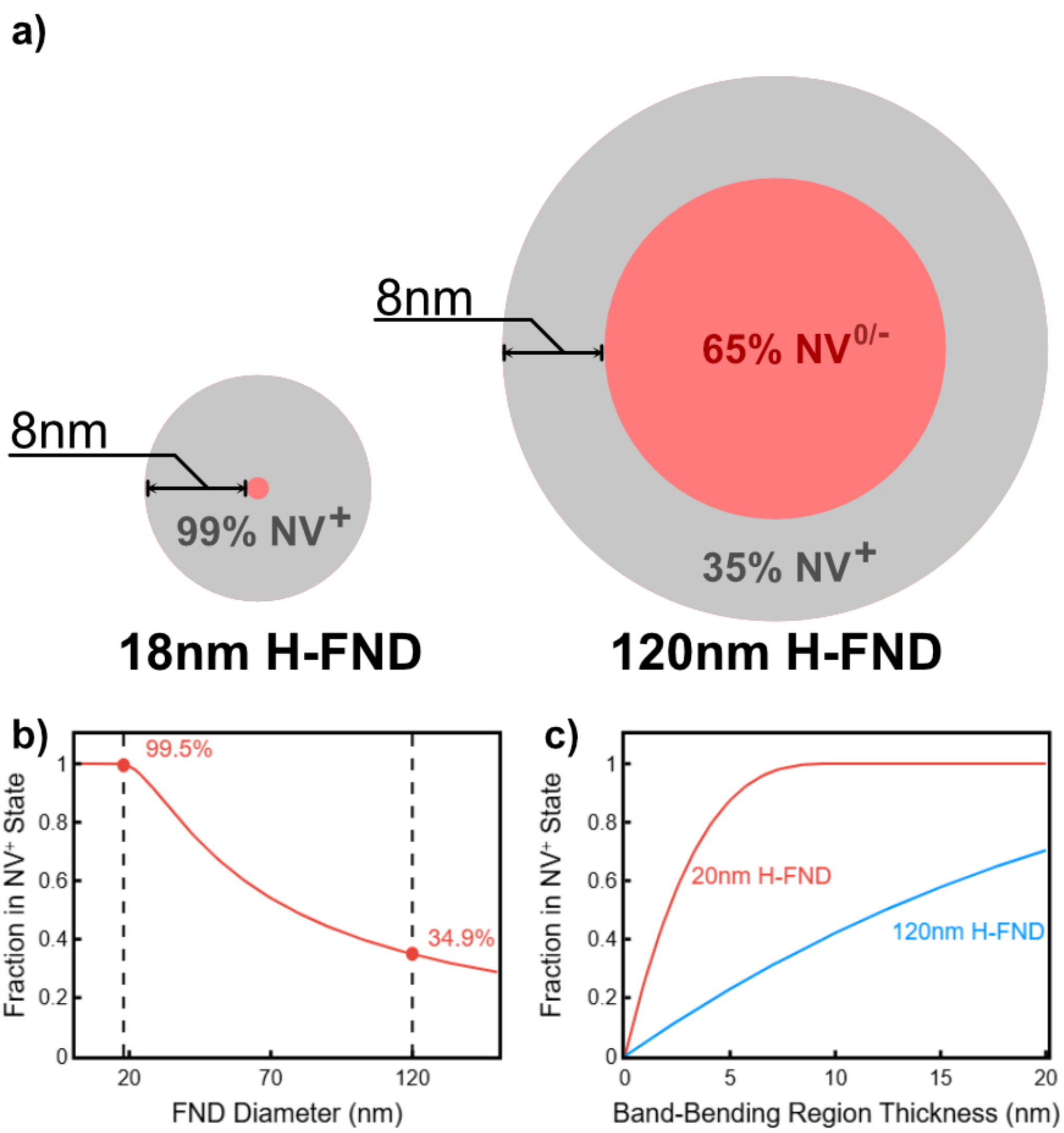

**FIGURE S6 | a)** Schematic illustration of a simple model for the NV charge state distribution inside FNDs, assuming a spherical particle shape. The model assumes that all NVs less than 8 nm from the surface are in the $NV^+$ charge state, and those further from the surface are in the $NV^0$ charge state. b-c) Fraction of NV centers in the $NV^+$ charge state as a function of FND particle diameter (b) and as a function of the band-bending region thickness (8 nm in panels a and b) for 20 nm (red trace) and 120 nm (blue trace) FNDs (c).

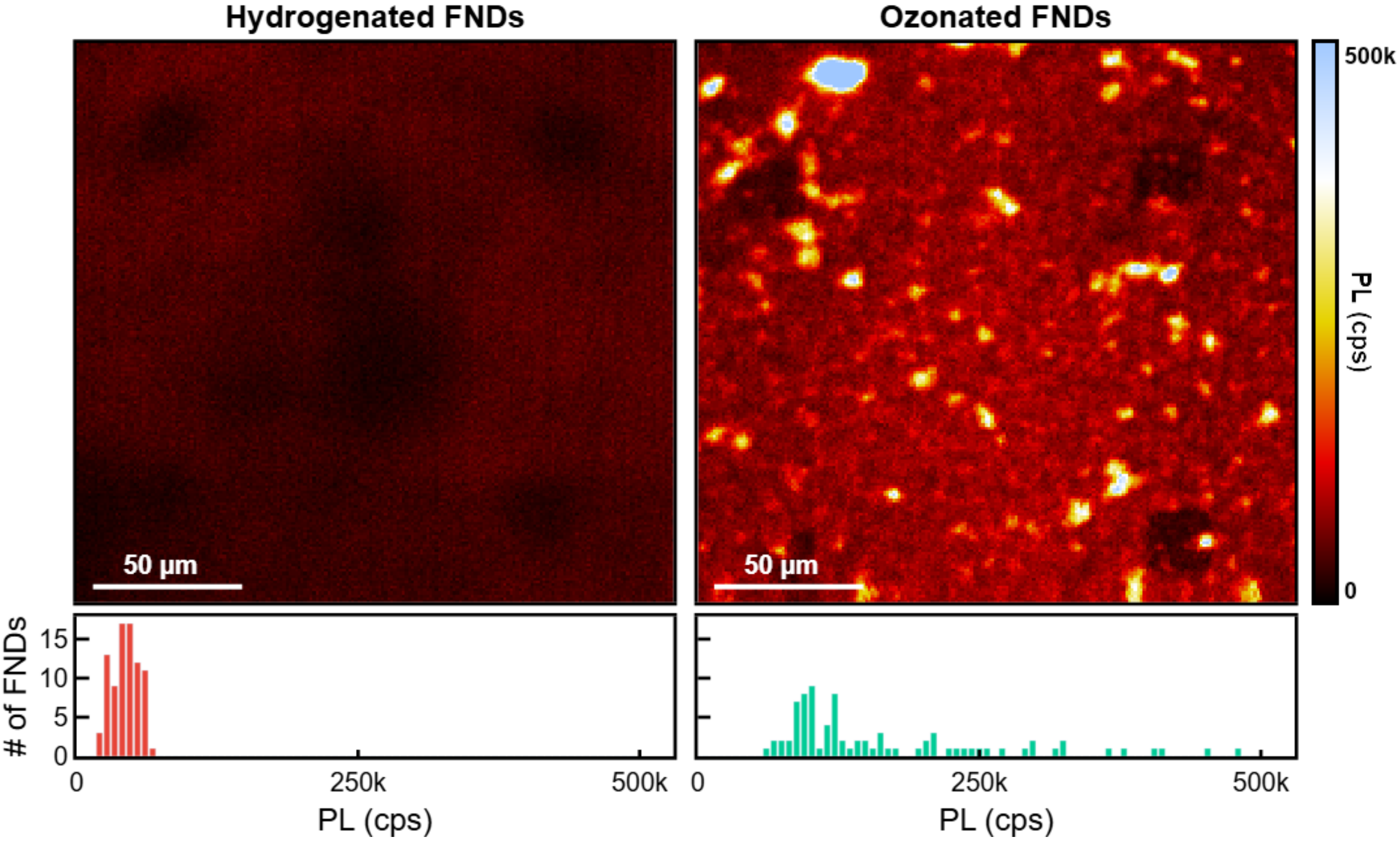

**FIGURE S7 |** Confocal fluorescence images of the same region of 18 nm FND particles distributed on a platinum-marked quartz substrate (see S1) acquired using a custom-built confocal microscope (see Figure S2) before (left, hydrogenated) and after (right, ozonated) 15 minutes in a benchtop UV-ozone cleaner. Both images are plotted using the same dynamic range. Below the images are corresponding histograms of the FND particle brightness. All features with a PL brightness above 60 kcps were identified as FNDs in the image on the right, and both images were aligned using the platinum markers to enable a direct comparison of the FND brightness in the histograms. This data is also shown in a box plot in the main text Figure 2d).

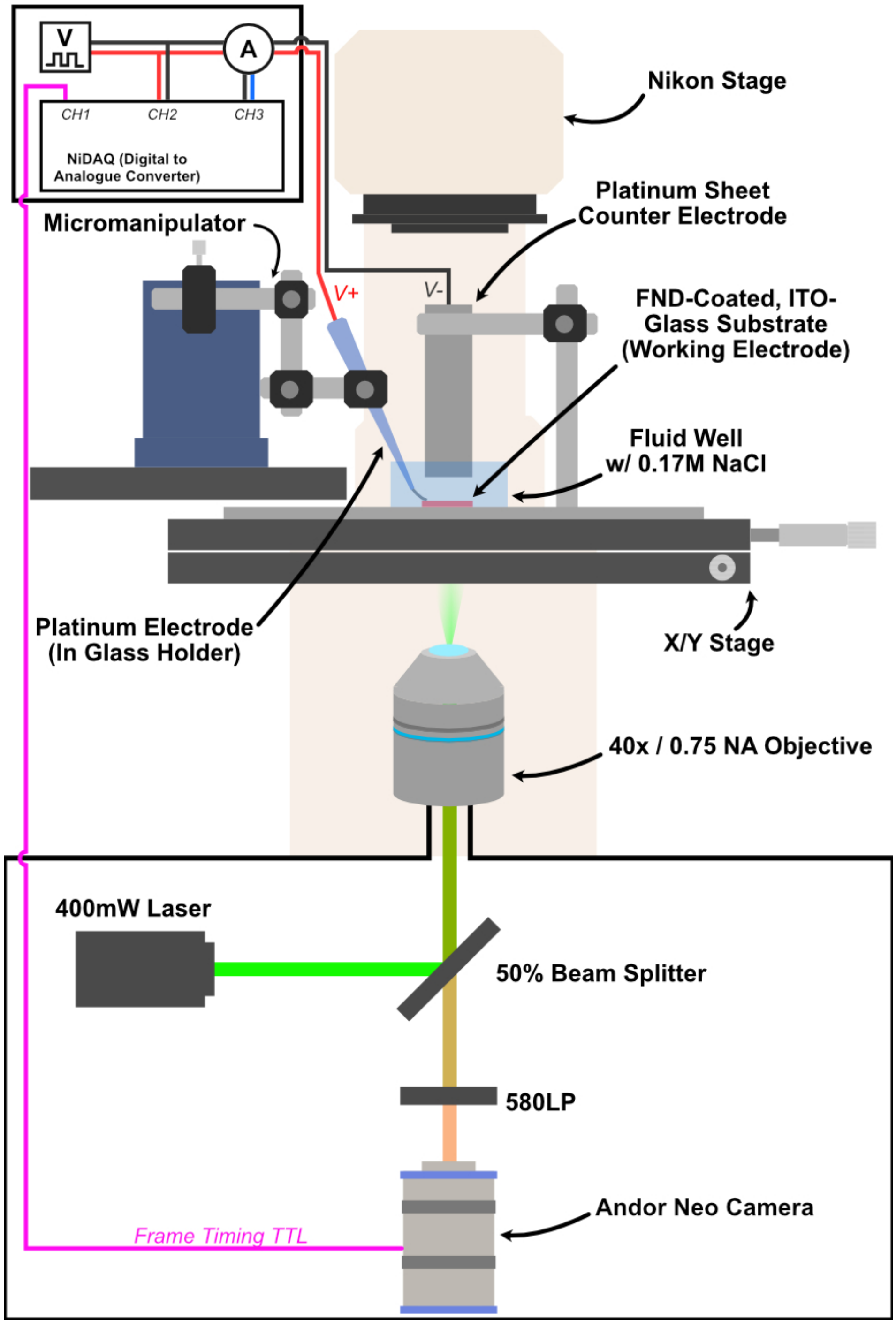

**FIGURE S8** | Diagram of the custom widefield fluorescence microscope used for voltage imaging experiments. The microscope is built on a "Nikon Eclipse" stage and uses an Andor Neo Zyla camera for high-speed video capture. Simultaneously, voltage, current, and TTL frame timing pulses are captured on a NiDAQ analogue-to-digital converter. The sample is excited by a 2W 532 nm CW laser through a 50% beam-splitter, and PL is collected through a 580 nm long-pass filter. The FND-coated ITO-glass substrate is submerged in 0.17 M NaCl solution inside a glass-bottomed microscope fluid-well and held against the X/Y stage using a custom 3D-printed ITO and platinum counter-electrode, which are connected via a platinum wire to the voltage generator and measurement circuitry.

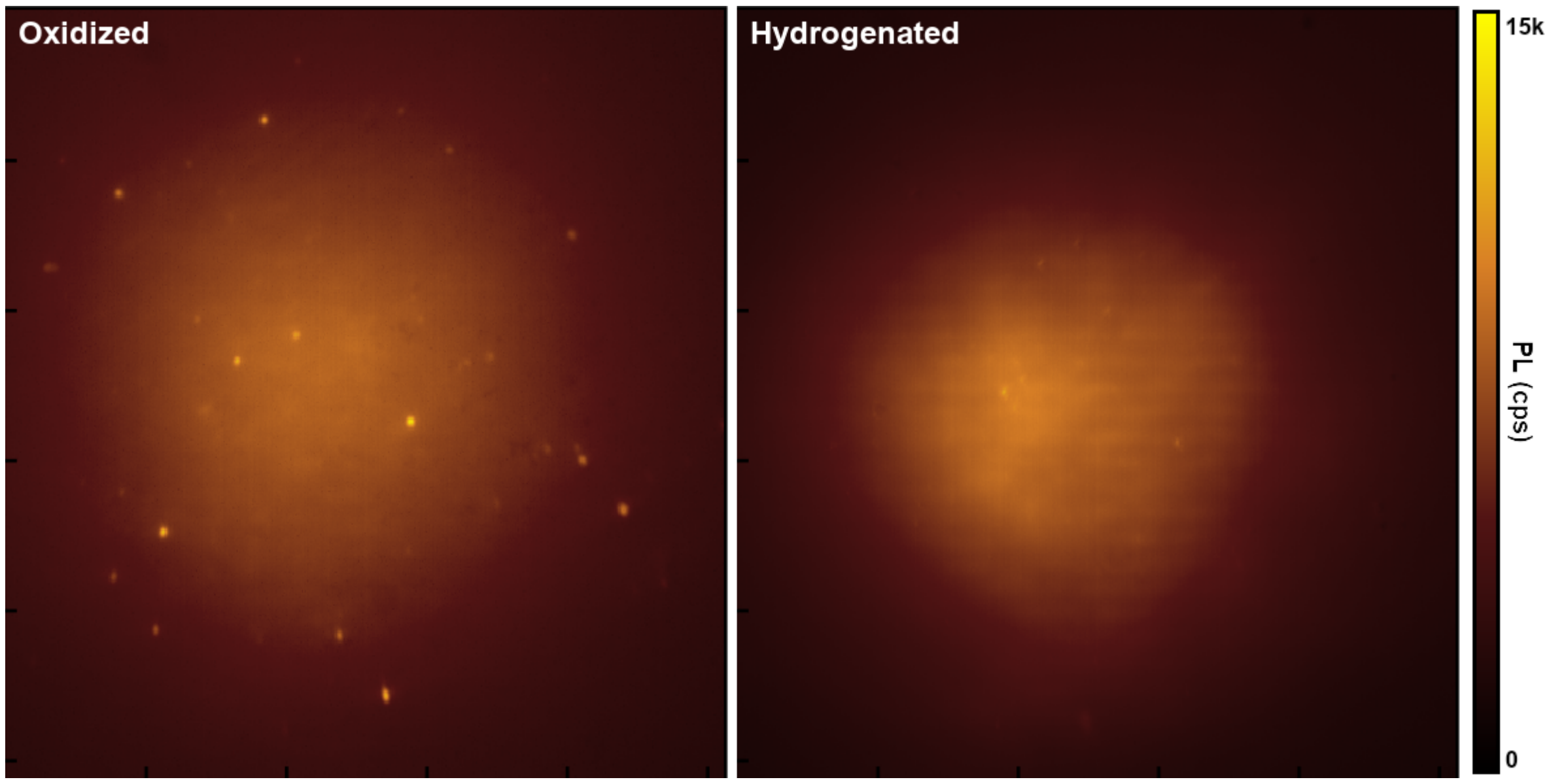

**FIGURE S9** | First frame of voltage-sensing experiments analysed in Figure 4, main text, and SI Figure S13. **(left)** Fluorescence widefield image of 18 nm FND-Oxy particles spin-coated on ITO-glass with 0 V applied. **(right)** Fluorescence widefield image of 18 nm FND-Hyd particles spin-coated on ITO-glass with 0 V applied. Both images use the same color scale shown on the right and have undergone the same background-removal processing. Hydrogenated particles/aggregates have lower base PL intensities than their oxidized counterparts.

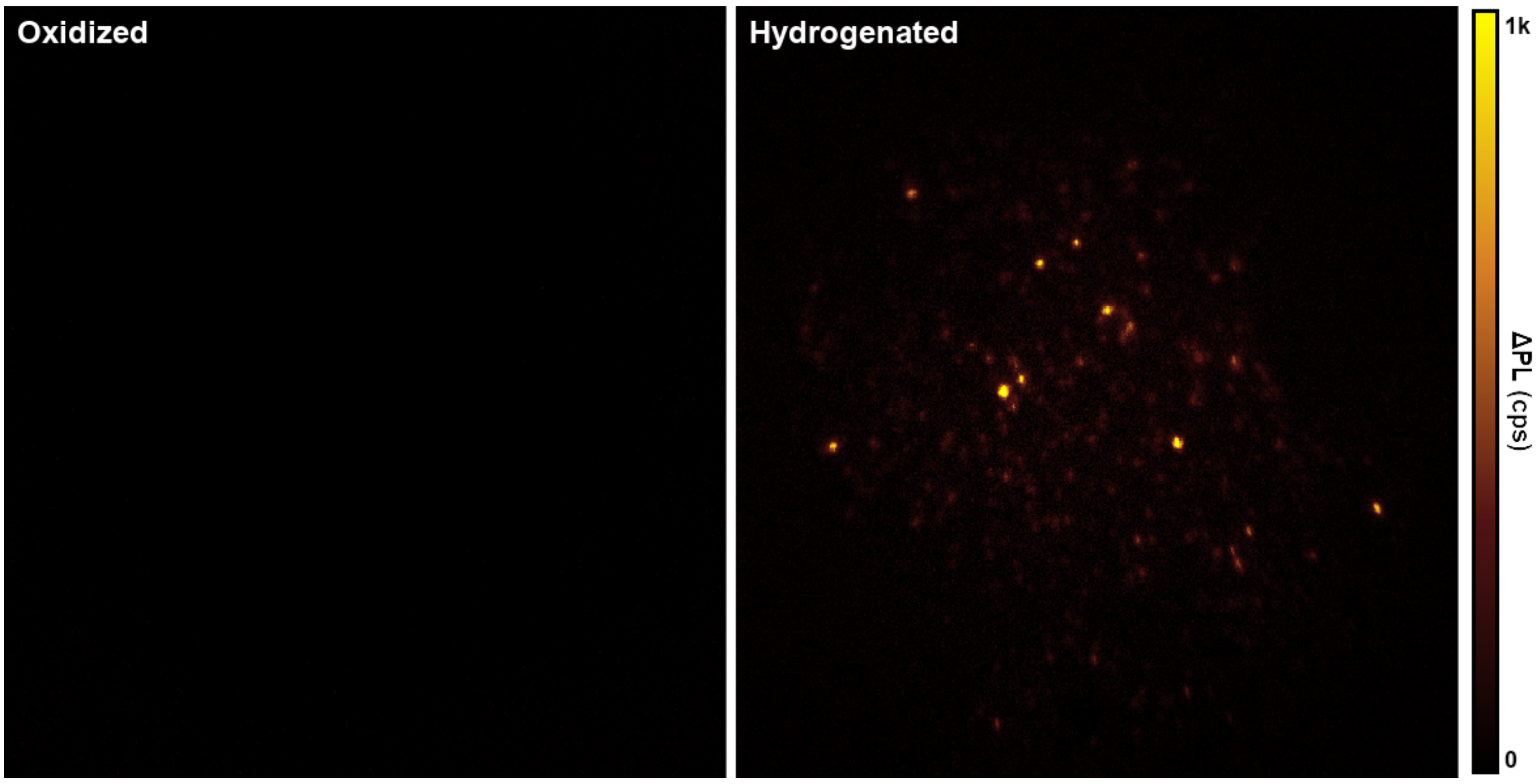

**FIGURE S10** | Change in PL (cps) between 0 and -1 V of FND-Oxy (left) and FND-Hyd (right) particles spin-coated onto an ITO-coated glass substrate. Same data as used in Figure 4b and Figure S5. Whilst FND-Oxy particles have a high raw PL, making them readily visible in Figure S10 (left), they exhibit much lower voltage sensitivity. Conversely, FND-Hyd particles are almost not visible in Figure S10, right, because of their low absolute PL intensity, but are visible here due to their high voltage sensitivity.

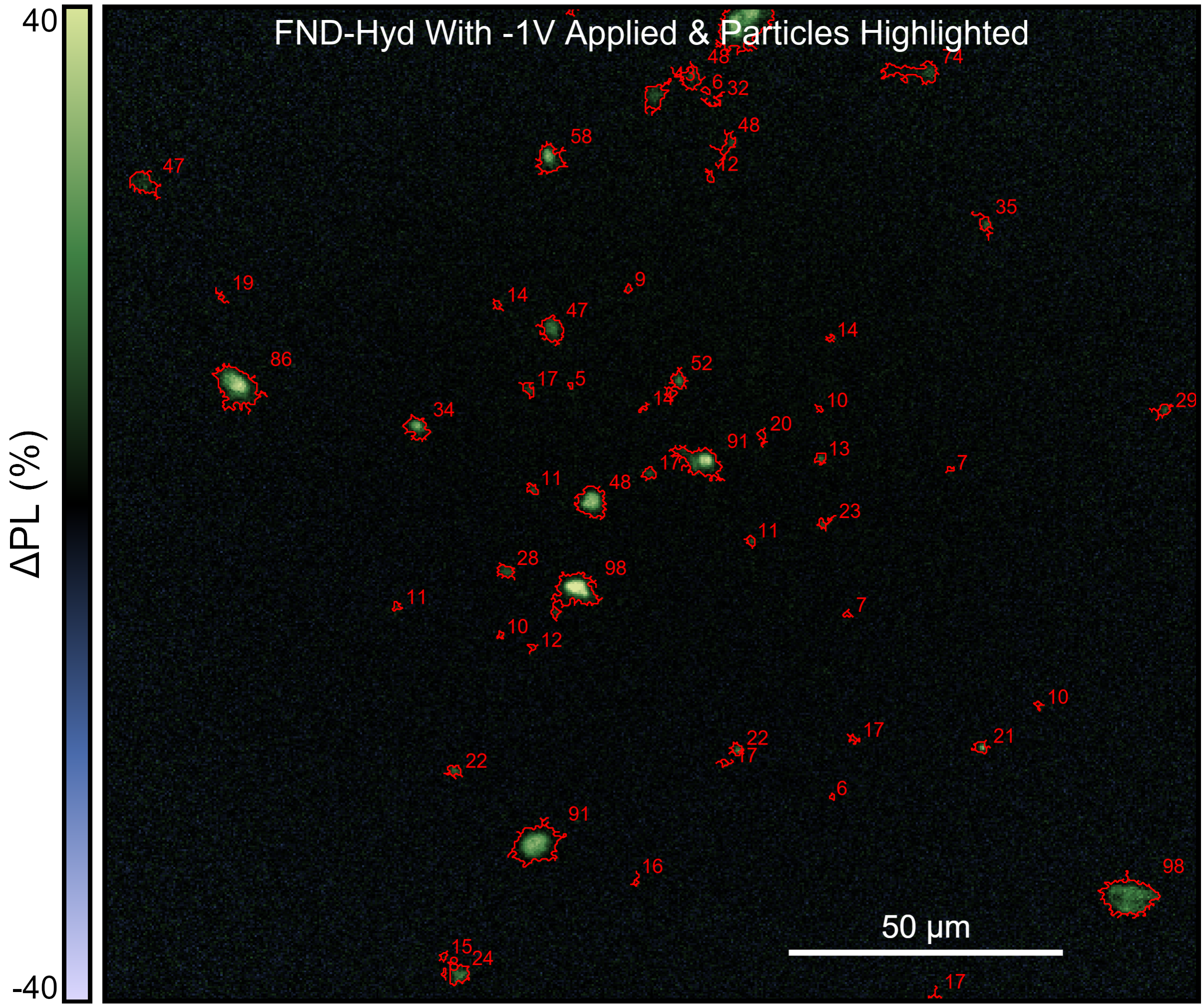

**FIGURE S11** | Widefield fluorescence image of *ΔPL* of FND-Hyd particles spin-coated on an ITO-glass substrate, used in the statistical analysis in the main text Figures 4h-i. Particles are identified using a threshold of *ΔPL* = 1% to identify particles in the image. The contour-detection feature of OpenCV (https://opencv.org/) was used to identify and separate individual particles and small aggregates. The index of each analysed particle is displayed next to each particle in the image.

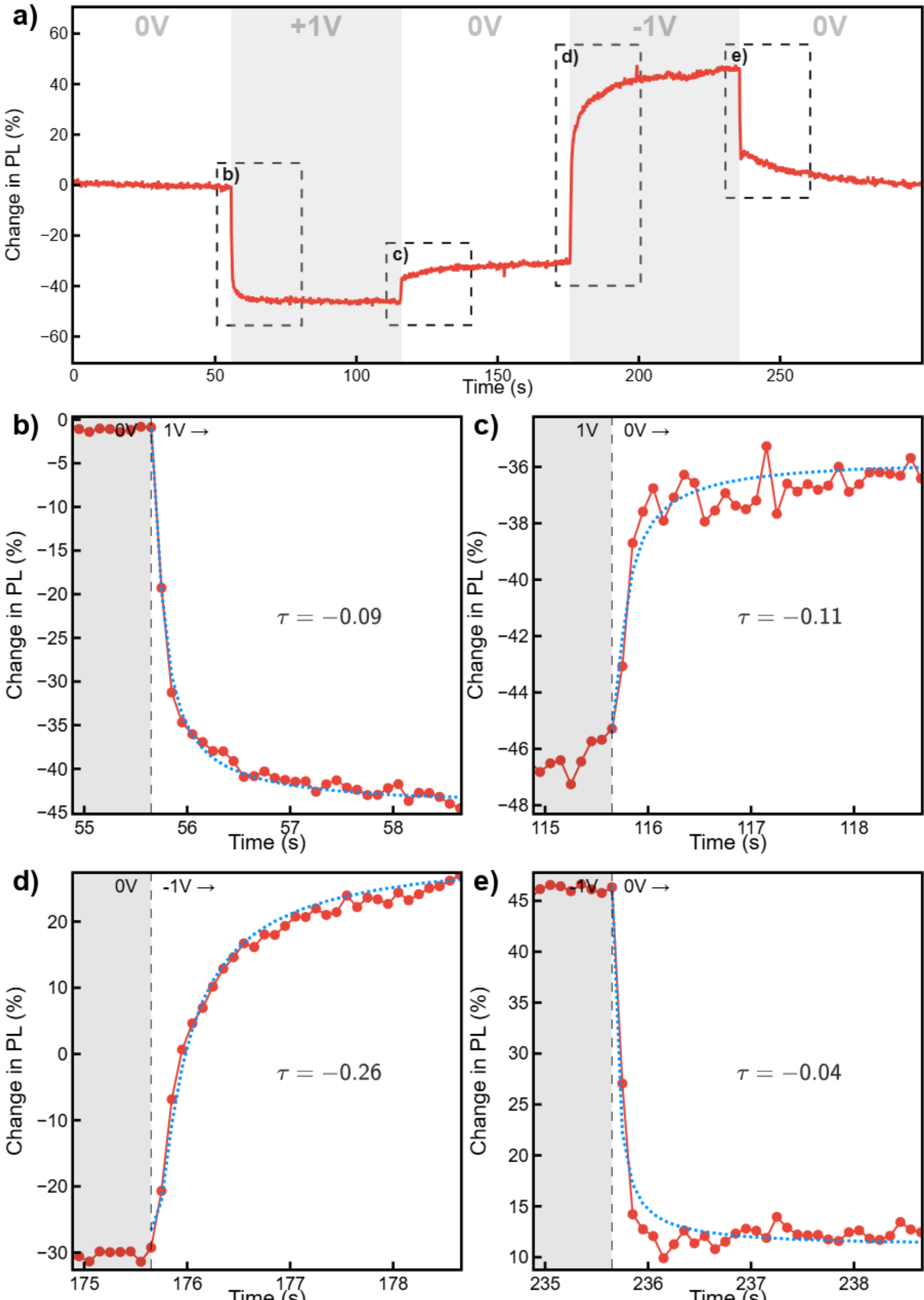

**FIGURE S12** | **a)** ΔPL as a function of time for a single FND-Hyd aggregate on an ITO-glass substrate measured using a custom widefield fluorescence microscope with 532 nm laser excitation and PL collected above 580 nm at 10 frames per second, as discrete voltages (+1V, -1V, 0V) are applied. **b-e)** ΔPL over time dynamics during voltage transitions at 10 FPS plotted against an exponential fit ($y = a + b * \exp(\tau/t)$). The characteristic time-constant ($\tau$) of each transition is displayed in each frame. Both **b)** and **d)** display a similar behaviour, with the time-constant of the exponential fit suggesting that the change in PL occurs ~3.4 times faster when switching to a positive voltage compared to a negative one. Both **c)** and **e)** also are like each other in that the bulk of their ΔPL transition occurs within approximately 2 frames (100 ms) after switching to 0 V.

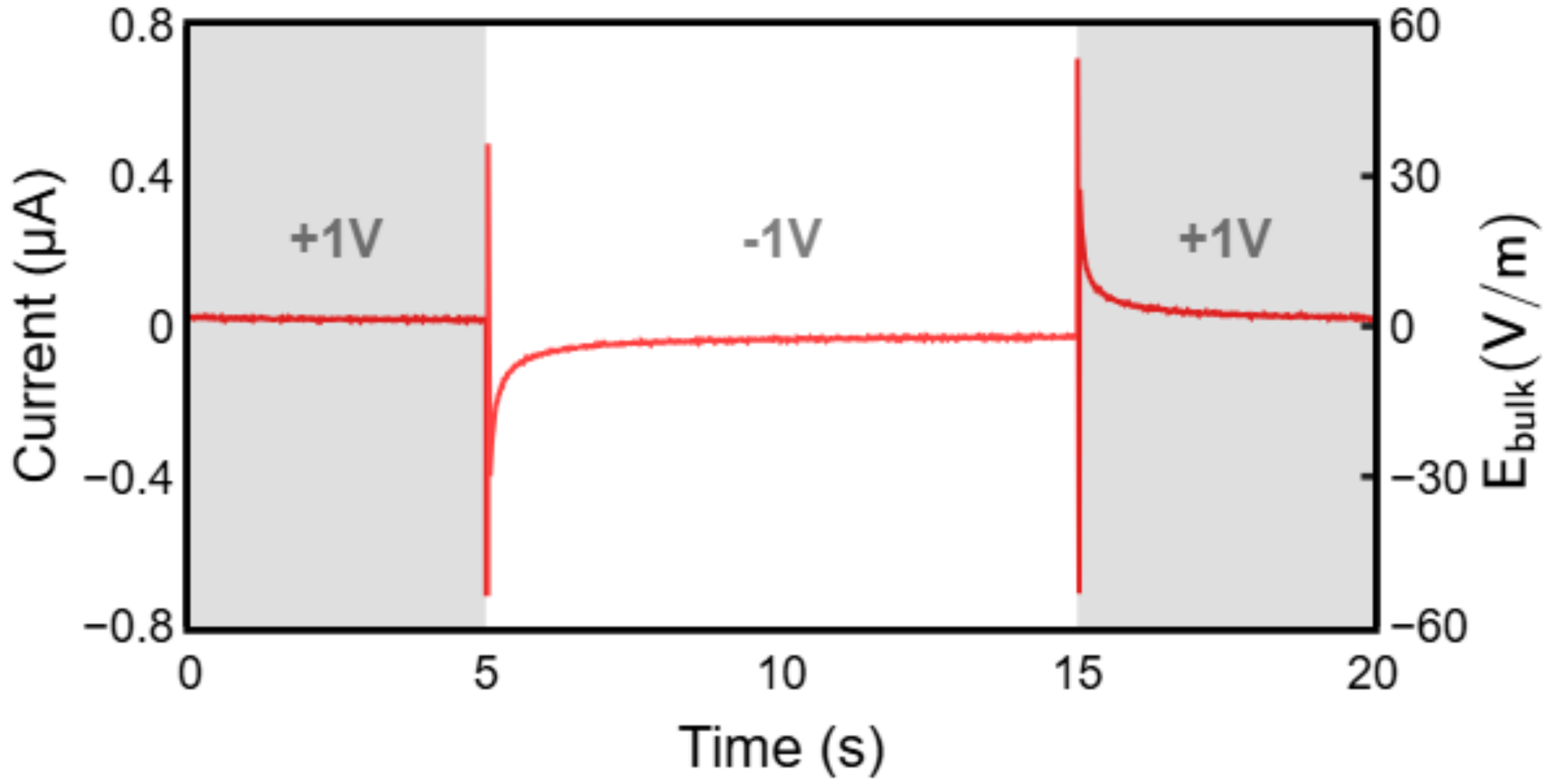

**FIGURE S13 | (left y-axis)** Electric current as a function of time during the measurements shown in Figure 5 in the main text as the applied voltage is switched from +1 V→ -1 V → +1 V. **(right y-axis)** Conversion of the current into residual electric field strength using the formula: $E_{bulk} = I/\kappa A$, where $I$ is the current in Amperes, $\kappa$ is the solution conductivity (15 S/m for 1.7M NaCl), and $A$ is the exposed surface area of the electrode. The current was measured using a Picoammeter (Keithley Instruments, USA) coupled into a digital-to-analogue converter (National Instruments, USA). When the voltage is applied, the electric field strength spikes above the measurement range of the picoammeter, then reduces to a small residual current as an electric double layer forms to shield the applied potential. Because the electric field strength in its steady state is small (~2 V $m^{-1}$) and uniform between the electrodes, it cannot explain the changes in PL observed in the main text Figure 5b.

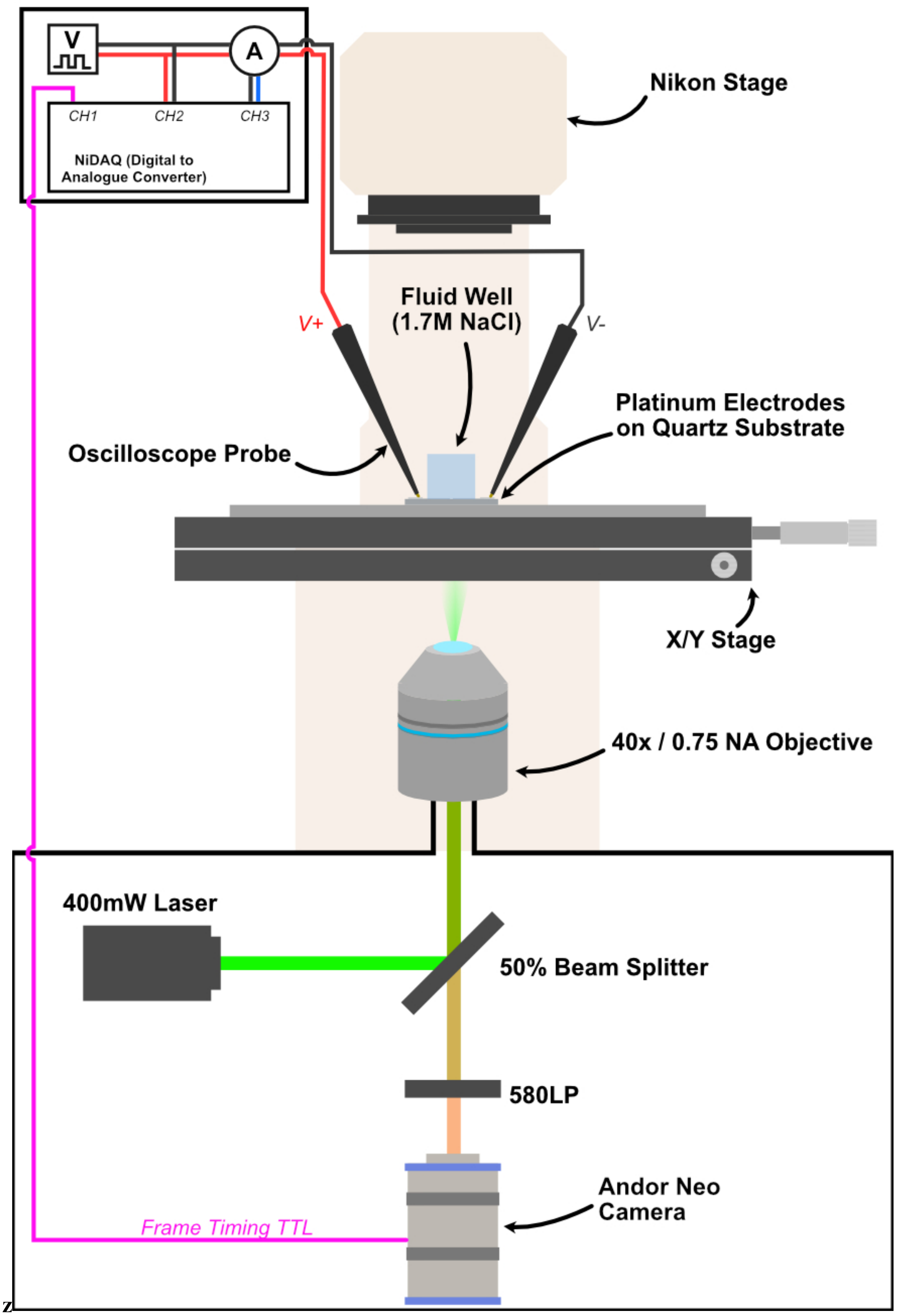

**FIGURE S14** | Diagram of the custom widefield fluorescence microscope used for ion-concentration imaging experiments in Figure 5 in the main text. The microscope is built on a "Nikon Eclipse" stage and uses an Andor Neo camera for high-speed video capture. Simultaneously, voltage, current and TTL frame timing pulses are captured on a NiDAQ analogue to digital converter. The sample is excited by a 4W laser through a 50% beam-splitter and captured after a 580 nm long-pass filter. Custom-built platinum electrodes are held in a custom 3D-printed holder above the X/Y stage, and electrodes are connected via oscilloscope probes to a voltage generator and measurement circuitry.

.

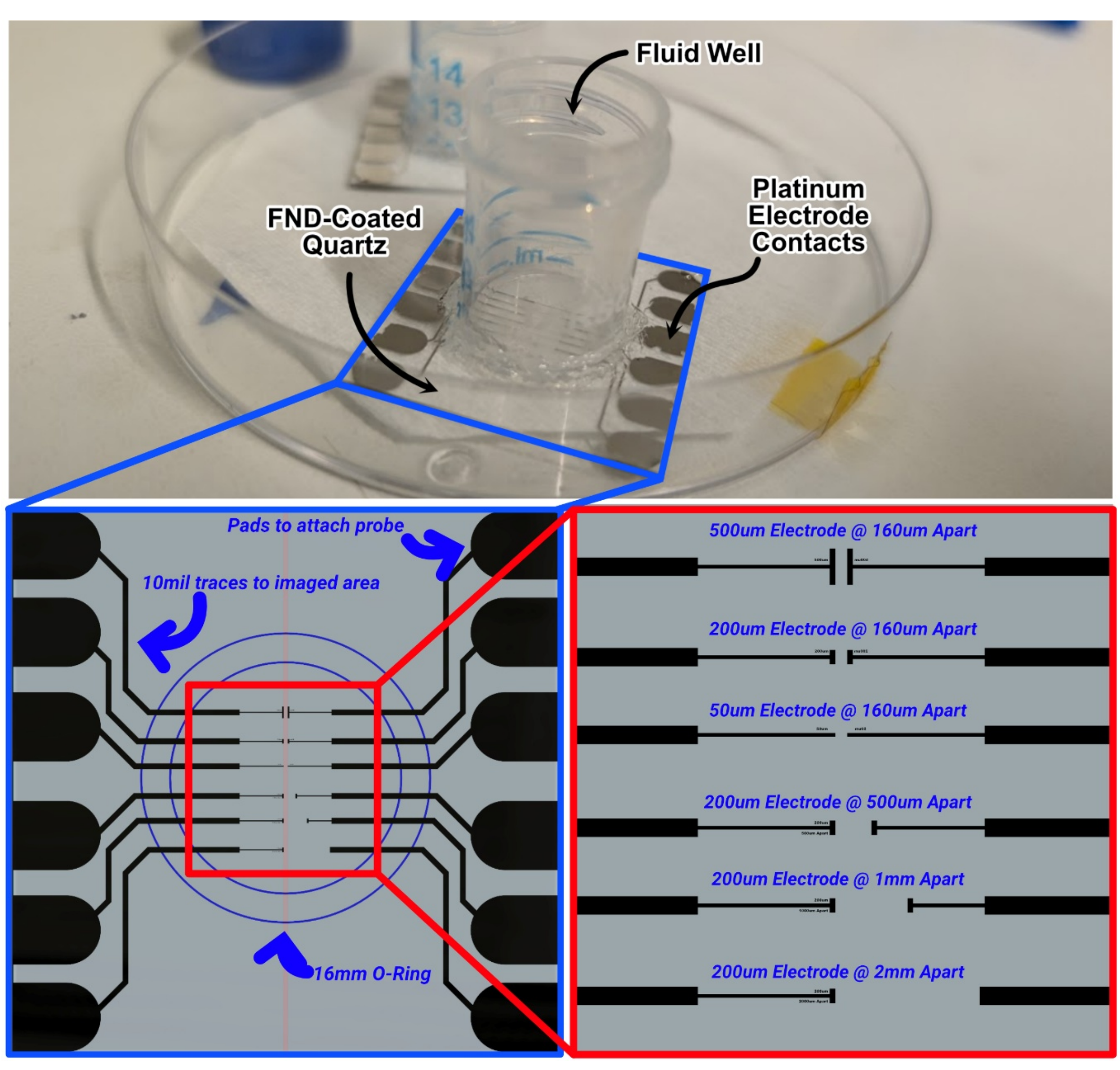

**FIGURE S15 |** Custom-made platinum electrodes and fluid-well used for the ion-concentration imaging experiments. A 1 × 32 × 32 mm fused silica (quartz) substrate was patterned using an MLA-150 maskless aligner (Heidelberg Instruments, Germany), and a 100 nm platinum layer was deposited using an electron-beam physical vapor deposition chamber (Kurt J. Lesker, Germany) and photoresist was removed via sonication in acetone. A centrifuge tube was then truncated and glued to the top surface, creating a fluid well with external access to electrode contacts. The electrodes used in experiments were the 500 μm wide / 160 μm apart.

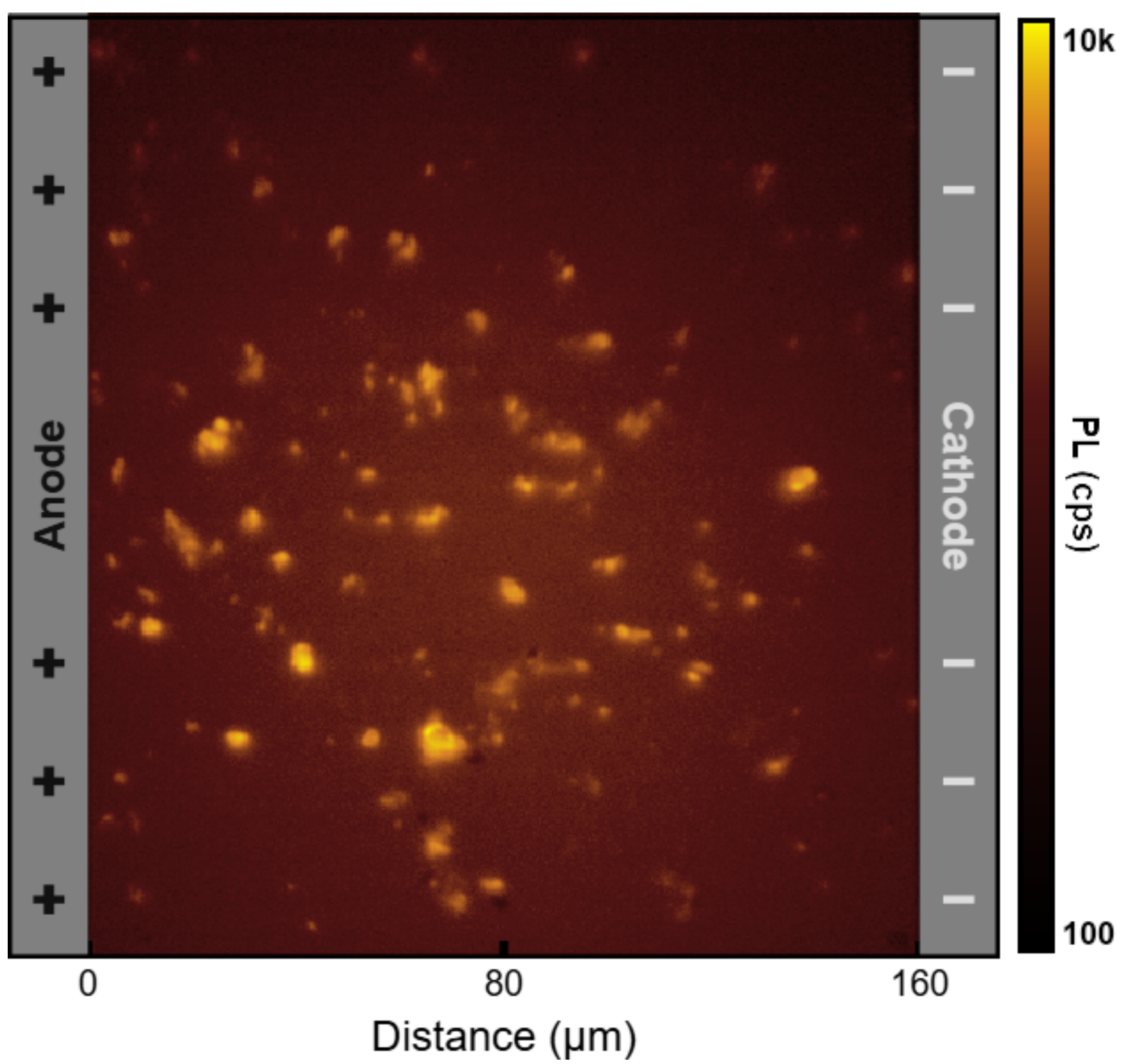

**FIGURE S16** | Widefield PL image of 18 nm FND-Hyd particles dispersed on a quartz substrate between platinum electrodes, illustrated in the main text Figure 5 a). The image was acquired using a custom-built widefield-fluorescence microscope, exciting at 532 nm and collecting PL above 580 nm. The particles are immersed in a 1.7 M NaCl solution, and 1 V is applied.

**VIDEO S17** | (See file "S18 2D Ion-Concentration Widefield PL Video.mp4") Video of 18 nm FND-Hyd ΔPL over time as the voltage is switched from +1 V to -1 V and back several times as indicated in the top left corner of the video using the setup shown in Figure 5 a) in the main text. A 1 Hz low-pass filter was applied. FND-Hyd particles are dispersed on a glass substrate between the electrodes, as shown in SI Figure S17. The experimental setup is described in Si Figures S15 and S16.

**VIDEO S18** | *(See file "S19 2D Ion-Concentration Widefield Concentration Video.mp4")* This video shows the same data shown in video S18, with the ΔPL values converted to changes in NaCl concentration based on our numerical simulation of the salt concentration over time.

**VIDEO S19** | *(See file "S20 - Voltage Sensing Experiment FND-Hyd Widefield Contrast Video")*. Video of 18 nm FND-Hyd ΔPL over time for FNDs dispersed on an ITO-glass substrate and submerged in 0.17M NaCl solution used as illustrated in main text Figure 4a) as the applied voltage is varied from -1 V to + 1 V. The voltage is applied to the ITO substrate with a platinum sheet counter-electrode suspended ~5 mm away from the ITO surface. The voltage is varied from -1 V to 1 V at a rate of -2 mV/s.

**All videos are available here:**

https://drive.google.com/drive/folders/1PYIRaSqKtedlfrCbFal94lqe9sjNX6Va?usp=sharing

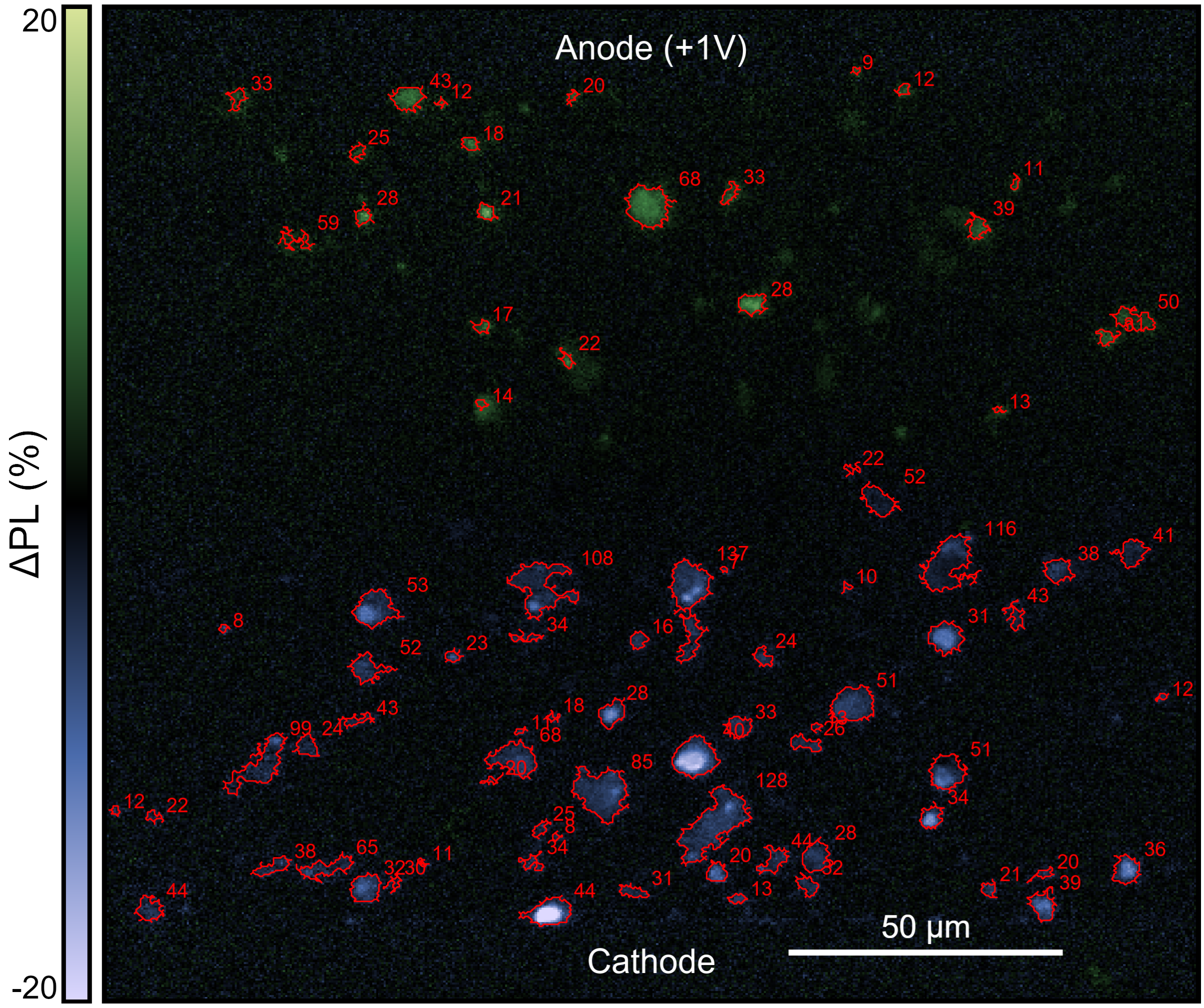

**FIGURE S20** | Widefield fluorescence image of the ΔPL of FND-Hyd particles self-assembled on a quartz substrate between two 500 µm wide platinum electrodes spaced 160 µm apart with 1 V applied between the anode (top) and cathode (bottom). This image is rotated by 90° relative to the image shown in the main text Figure 5b) and SI Figure S13. Particles are identified using a threshold of ΔPL > 1%. The contour-detection feature of OpenCV was used to identify and separate individual particles and small aggregates. The index of each analysed particle is displayed next to each particle in the image.

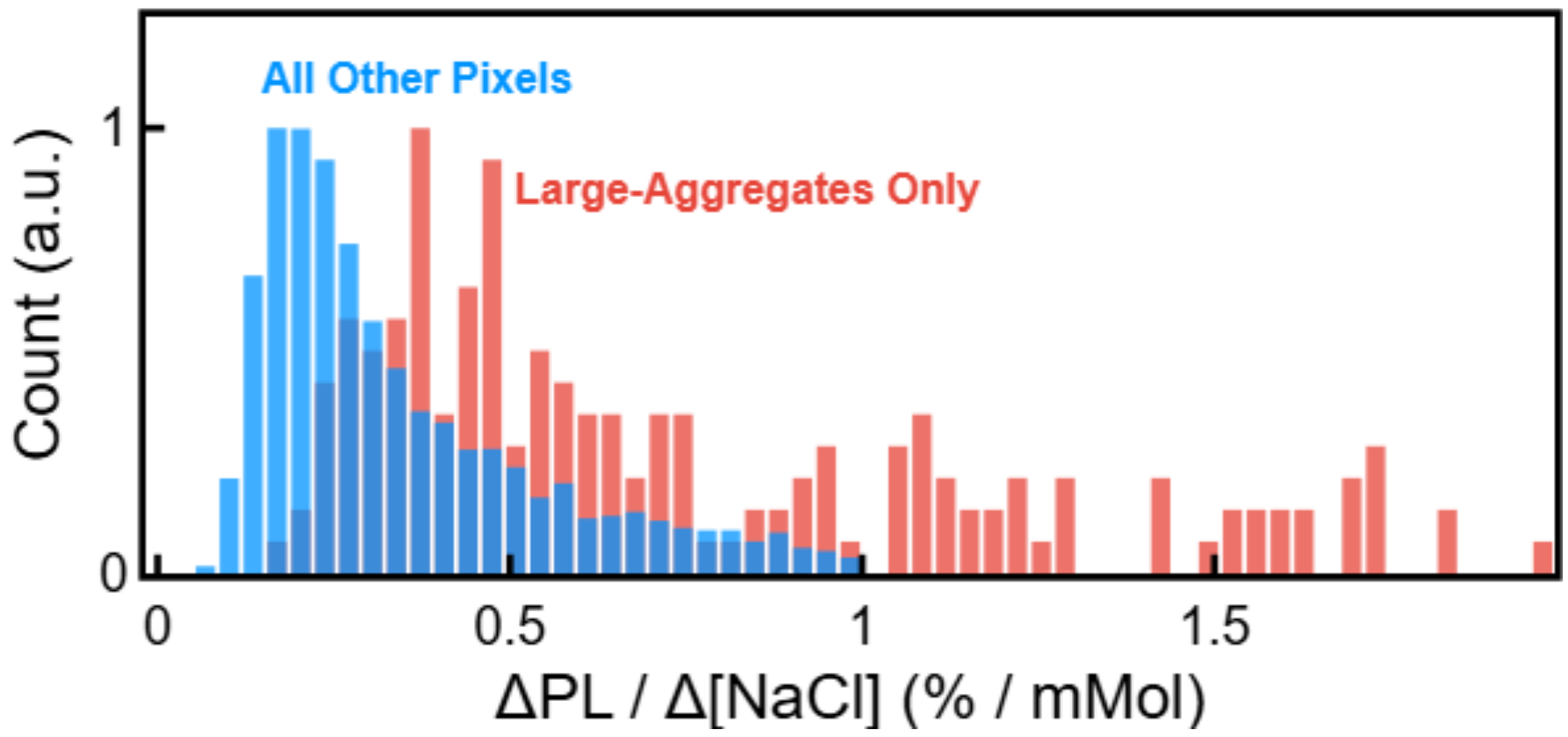

**FIGURE S21** | Comparison between the sensitivity ($\Delta PL/\Delta[NaCl]$ in $\%.mMol^{-1}$) of aggregates versus all pixels when -1 V is applied in the electrochemical cell using the particles identified in Figure S14 during ion-concentration imaging experiments. Pixels within the red contours of Figure S21 are identified as "large aggregates". The maximum ΔPL of each aggregate and pixel when -1 V is applied is measured and divided by the simulated NaCl concentration at the y-position of the pixel. The result indicates that non-aggregates still maintain a mean sensitivity of 0.29 $\%.\mathrm{mM}^{-1}$, which is less than half of the aggregates' mean sensitivity of 1.3 $\%\ \mathrm{mM}^{-1}$.

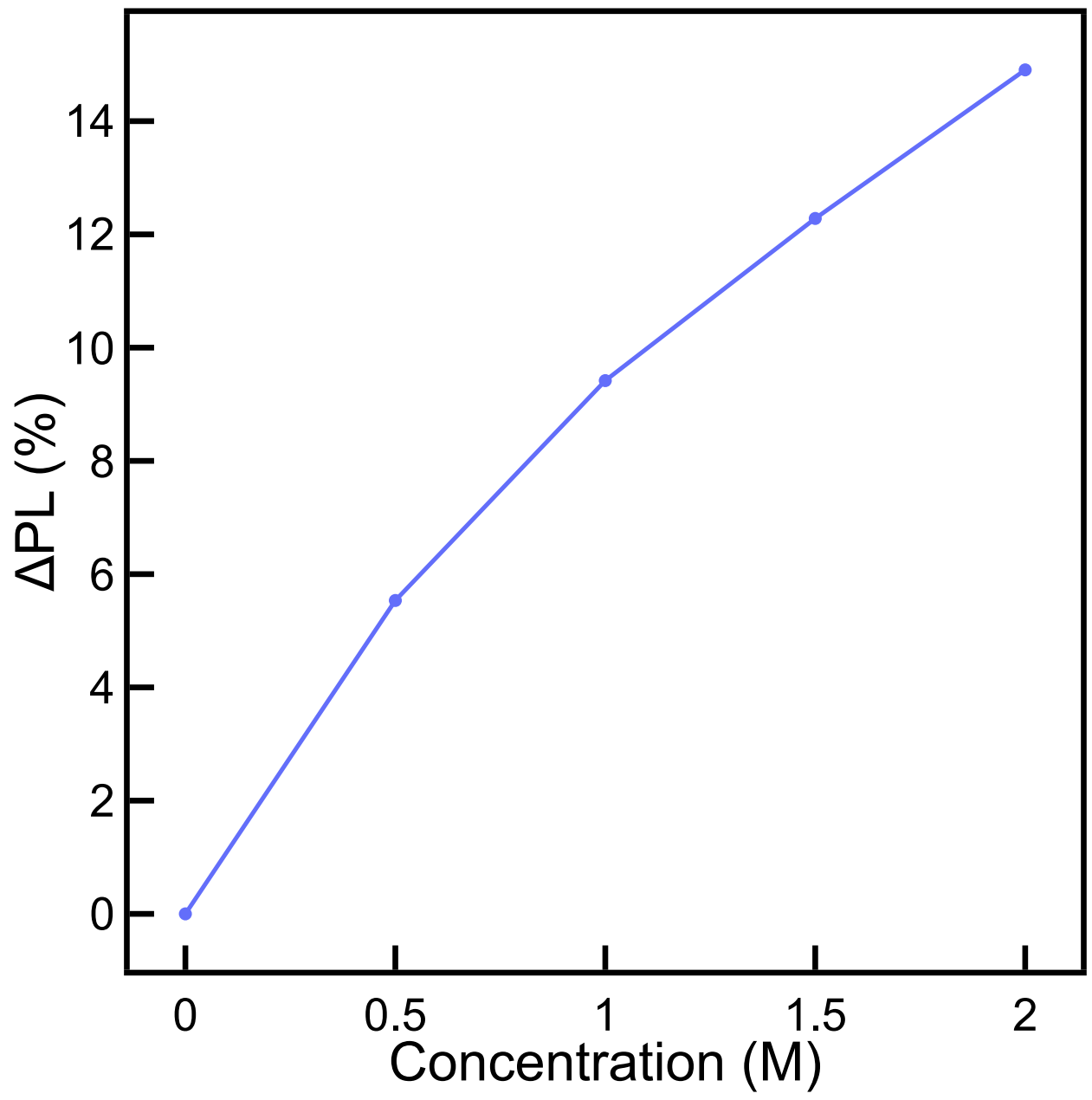

**FIGURE S22** | Average ΔPL (%) as a function of NaCl concentration for 18 nm FND-Hyd particles self-assembled on quartz as NaCl is increased in discrete steps of 0.5M by adding 3M NaCl solution to DI water. Measurements were acquired using a custom widefield fluorescence microscope and ΔPL values calculated based on a 400 × 400 pixel region in the centre of the frame using the equation $\Delta PL=(I-I_0)/I_0$, where $I$ and $I_0$ are the average PL intensities in the 400 × 400 pixel region in the presence of NaCl and no NaCl added, respectively, except at [NaCl] = 0, where $I = I_0$.

**SECTION S23** | ION CONCENTRATION GRADIENT MODELLING

In figures 3,4 and 5, simulations of ion concentration gradients due to electromigration were used to help correlate and justify the ion-concentration dependence of the observed changes in PL intensity in H-FNDs. To achieve this, the experiment was modelled using a Nernst-Planck model of electromigration and recreated for simulation. These simulations were created using the tertiary-current-distribution (tcd) interface in the electrochemistry module of COMSOL Multiphysics 6.2. Also provided is a detailed explanation of the key concepts behind this modelling.

The generation of bulk ion concentration gradients is often misunderstood in electrochemistry because it is often though that when a voltage is applied so a solution, that ions in the solution move towards their oppositely charged electrode. However, for low voltages under blocking conditions (no Faradaic reactions), and outside of the electric-double-layer, this is not the case. Instead, the electroneutrality principle states that ions will move in tandem towards an electrode such that no local electric fields are generated by charge separation. Mathematically, the electroneutrality condition is given by equation 1.

$$\sum z_i c_i = 0 \tag{1}$$

Where $z_i$ is the species charge of ion $i$, and $c_i$ is its concentration. Coulomb forces between ions ensure macroscopic electroneutrality such that any changes in the accumulation of space charge are compensated by electric field-driven ion movement[4,5]. The only place where this does not hold true is within a Debye length of the electrode where non-electroneutral double layers form which shield the potential difference between electrodes[4,5]. In our dilute 1:1 bulk solution, electroneutrality is assumed in all places except for within a Debye length of the electrode. What this means for this experiment is that by applying a voltage to our salt solution, we are seeing the change in concentration of both $c_{Na}$ and $c_{Cl}$ together, resulting in a change in the bulk concentration $c_{bulk}$ as opposed to the concentration of any one individual ion.

$$J_i(x,t) = \underbrace{-D_i \nabla c_i}_{Diffusion} + \underbrace{\frac{D_i z_i e}{k_B T} c_i \boldsymbol{E}}_{Migration} + \underbrace{c_i \boldsymbol{v}}_{Advection} \tag{2}$$

The above Nernst-Planck model of electrochemical mass transport explains why this occurs. In essence, this states that when a voltage is applied, the total flux $J_i$ of a given species $i$ is equal to the sum of the diffusion, migration and advection terms, where $D_i$ is the diffusivity of the chemical species, $\nabla c_i$ is the concentration gradient, $e$ is the elementary charge, $k_B$ is the Boltzmann constant, $T$ is the absolute temperature and $\boldsymbol{E}$ is the electric field vector. In our experimental setup we can assume that no advection takes place, thus removing the advection term, and that the electric field is static and constant so the vector $\boldsymbol{E}$ becomes the electric field strength $E$ in units $V.m^{-1}$. By removing the time dependency, the concentration gradient over time and space also simplifies it to the concentration over just space $\nabla c_i \rightarrow \Delta c_i$. We can also simplify the constant terms to $R$ such that:

$$R = \frac{eE}{K_B T} \tag{3}$$

Which simplifies the equation to just:

$$J_i = D_i(z_i c_i R - \Delta c_i) \tag{4}$$

Here, we see that the flux of an ion is proportional to the flux induced by the migration term, which moves the ion towards the oppositely charged electrode, minus the flux induced by the diffusion term, which moves the ion down its concentration gradient. To make a point, let's quickly imagine a system of dissolved NaCl. Without a voltage applied, the concentration of the ions is the same everywhere, and the concentration gradient is zero ($\Delta c_i = 0$). This means that upon applying a voltage, the initial ion flux is driven entirely by the migration term. As the system evolves, migration increases the concentration of the ion on one side compared to the other, thus, the $\Delta c_i$ term also increases which in turn counteracts the migration. The system finally reaches a steady state when the diffusion term matches the magnitude of the migration term:

$$\underbrace{J_i}_{Flux} = 0 \quad when \quad \underbrace{z_i c_i R}_{migration} = \underbrace{\Delta c_i}_{Diffusion} \tag{5}$$

For this to occur, the system must then have a non-zero concentration gradient, thus, more of one species of ion on one side than the other. However, there remains the question of why do *both* ions move in one direction, and which ions choose which direction the ions move? The answer lies in the weeds of the details. If we look at the different ions in our practical example, we find that in aqueous solutions they each have a different coefficient of diffusion, the values of which are shown below[6,7].

$$D_{Na} = 1.33 \cdot 10^{-9}\ [m^2.s^{-1}], \qquad D_{Cl} = 2.03 \cdot 10^{-9}\ \ [m^2.s^{-1}] \tag{6}$$

Because the flux of ion is directly proportional to the diffusion coefficient ($J_i \propto D_i$):

$$D_{Cl} > D_{Na} \tag{7}$$

$$J_i \propto D_i \tag{8}$$

$$J_{Cl} > J_{Na} \tag{9}$$

This means that the flux of $Cl$ is greater and thus will be pulled more strongly than $Na$ by the anode, such that the total flux $\boldsymbol{J} = \sum J_i$ is given by:

$$\boldsymbol{J} = J_{Na} - J_{Cl} \tag{10}$$

$$\therefore \boldsymbol{J} < 0 \tag{11}$$

This means that for the exact same ion concentration, concentration gradient and applied electric potential, the flux of $Cl$ is going to be stronger than the flux of $Na$, meaning the dominant flux is $J_{Cl}$ and both ions will move towards the anode. To make this more intuitive, imagine a similar scenario to before, but in which we have only a single $Na$ and $Cl$ ion. At the time just after we apply the voltage ($t \to 0^+$), the anode attracts $Cl$ ion more strongly than the cathode attract the $Na$ ion causing a macroscopic charge separation, which generates a small Coulombic force, attracting the $Na$ ion to towards the $Cl$ ion, and moving both in the direction of the anode. This is the premise behind electroneutrality. The bulk ion concentration over the entire electro chemical cell changes, but the local concentrations of each ion remain identical.

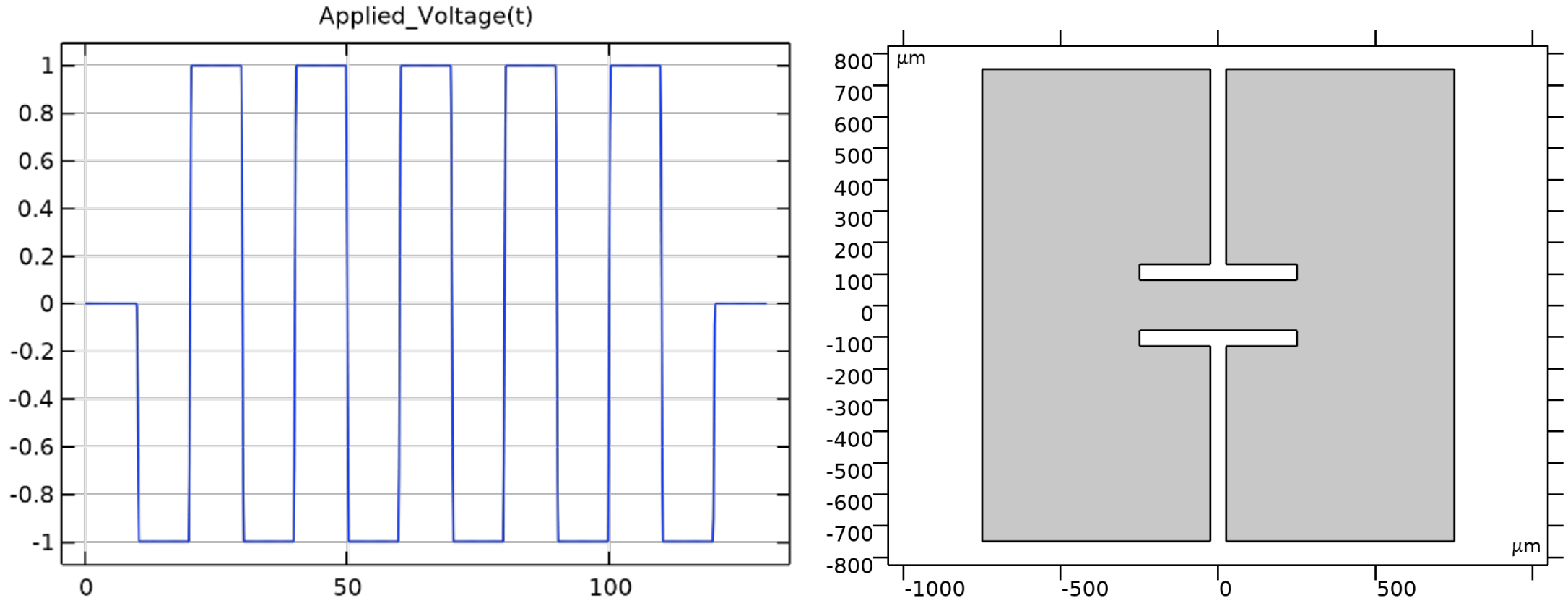

**FIGURE 23.1 | Left)** Applied voltage over time square wave using in an iteration of the ion-concentration sensing experiment simulations. **Right)** 2D geometry of the ion-concentration sensing experiment. Bottom electrode was configured to exhibit the applied voltage in left), and top electrode configured as 0 V. Border was configured as an open boundary with a concentration of $c_0$ external to the simulation domain.

In COMSOL Multiphysics, time-dependent finite-element-analysis simulations using the tcd interface were created. These used the Nernst-Planck equation with the electroneutrality condition as described above. For the voltage sensing experiment, the experimental setup was replicated in 3-Dimensions due to the more complex geometry required. Here, the ITO-electrode was set up as an electrode with an applied square-wave voltage (matching the real-world waveform, +1 to -1 V), and the suspended counter electrode set at a constant 0 V. The electrolyte was defined as having diffusion coefficients of $D_{Na}$ and $D_{Cl}$ and a relative permittivity of 80.

In the ion-concentration sensing experiments, the dual-electrode geometry was modelled in 2D due to its simplicity. 3D models were also tested but found to provide the same result at the 2D counterpart whilst taking much longer to generate. Simulation parameters were defined in the same way as the voltage-sensing simulations; however, because the electrode geometry is extremely small, the domain of the experiment was reduced to just the relevant area, and outer boundaries replaced with an open-boundary condition, with a fixed external concentration of the starting bulk concentration $c_0 = 1711\, mol.m^{-1}$.

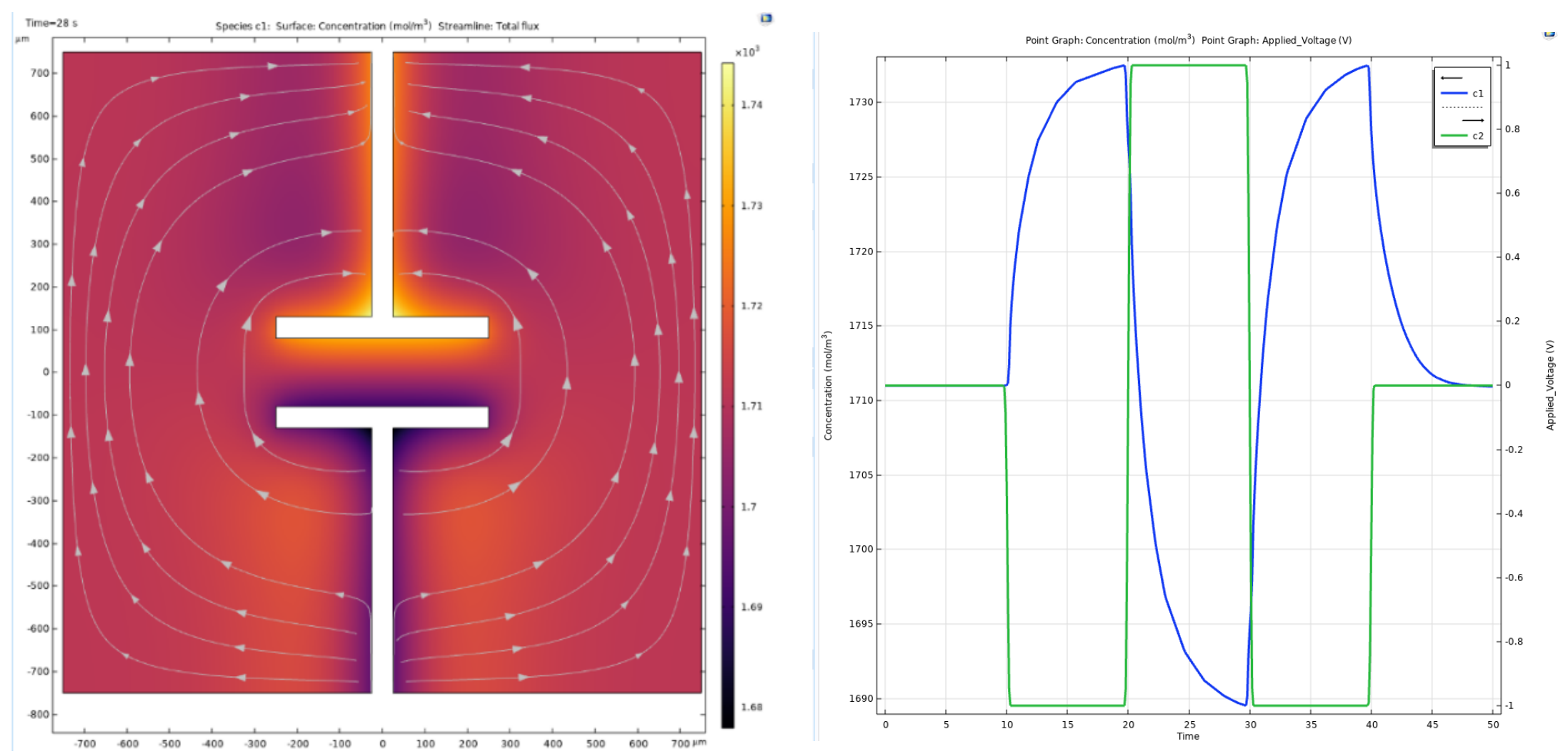

**FIGURE 23.2** | **(left)** 2D Ion concentration map of the simulation domain at time t=28s. **(right)** 1D plot of ion concentration (blue trace) and applied voltage (green trace) over time for a point 10µm away from the bottom electrode in (left).

The results of experiment provided absolute values for the concentration of ions in 2D over time, allowing us to view the movement of ions, and compare it to the movement of ions observed in figures 5c-e and SI figures 13 and 14. COMSOL's automatically generated report is also provided as a separate document.

**FILE S24** | *(See file "S24 – 2D Time-Dependent TCD Report.pdf")* Detailed report of the 2D time-dependent tertiary current distribution simulation generated by COMSOL Multiphysics is available here:

https://drive.google.com/drive/folders/1PYIRaSqKtedlfrCbFal94lqe9sjNX6Va?usp=sharing

## REFERENCES

[1] O. Chum, T. Pajdla, P. Sturm, *Comput. Vis. Image Underst.* **2005**, *97*, 86.

[2] "OpenCV: Basic concepts of the homography explained with code," can be found under https://docs.opencv.org/4.x/d9/dab/tutorial_homography.html, **n.d.**

[3] V. Petráková, A. Taylor, I. Kratochvílová, F. Fendrych, J. Vacík, J. Kučka, J. Štursa, P. Cígler, M. Ledvina, A. Fišerová, P. Kneppo, M. Nesládek, *Adv. Funct. Mater.* **2012**, *22*, 812.

[4] B. Janotta, M. Schalenbach, H. Tempel, R.-A. Eichel, *Electrochimica Acta* **2024**, *508*, 145280.

[5] M. Schalenbach, B. Hecker, B. Schmid, Y. E. Durmus, H. Tempel, H. Kungl, R.-A. Eichel, *Electrochem. Sci. Adv.* **2023**, *3*, e2100189.

[6] A. Marecka-Migacz, P. T. Mitkowski, A. Nędzarek, J. Różański, W. Szaferski, *Membranes* **2020**, *10*, DOI 10.3390/membranes10090235.

[7] Y. A. Le Gouellec, M. Elimelech, *Environ. Eng. Sci.* **2002**, *19*, 387.